\documentstyle[aps,prb,twocolumn,epsf,floats]{revtex}
\begin{document}
\draft
\wideabs{
\title{Composite Fermions and quantum Hall systems:
       Role of the Coulomb pseudopotential}
\author{
   Arkadiusz W\'ojs}
\address{
   University of Tennessee, Knoxville, Tennessee 37996, USA\\
   and Wroclaw University of Technology, 50-370 Wroclaw, Poland}
\author{
   John J. Quinn}
\address{
   University of Tennessee, Knoxville, Tennessee 37996, USA}
\maketitle
\begin{abstract}
The mean field composite Fermion (CF) picture successfully predicts
angular momenta of multiplets forming the lowest energy band in 
fractional quantum Hall (FQH) systems.
This success cannot be attributed to a cancellation between Coulomb 
and Chern--Simons interactions beyond the mean field, because these 
interactions have totally different energy scales. 
Rather, it results from the behavior of the Coulomb pseudopotential 
$V(L)$ (pair energy as a function of pair angular momentum) in the 
lowest Landau level (LL).
The class of short range repulsive pseudopotentials is defined that 
lead to short range Laughlin like correlations in many body systems 
and to which the CF model can be applied.
These Laughlin correlations are described quantitatively using the 
formalism of fractional parentage.
The discussion is illustrated with an analysis of the energy spectra 
obtained in numerical diagonalization of up to eleven electrons in 
the lowest and excited LL's.
The qualitative difference in the behavior of $V(L)$ is shown to 
sometimes invalidate the mean field CF picture when applied to higher 
LL's.
For example, the $\nu={7\over3}$ state is not a Laughlin 
$\nu={1\over3}$ state in the first excited LL.
The analysis of the involved pseudopotentials also explains the success 
or failure of the CF picture when applied to other systems of charged 
Fermions with Coulomb repulsion, such as the Laughlin quasiparticles 
in the FQH hierarchy or charged excitons in an electron--hole plasma.
\end{abstract}
\pacs{71.10.Pm, 73.20.Dx, 73.40.Hm}
}

\section{Introduction}
\label{secI}

The discovery of the integer (von Klitzing, Dorda, and Pepper 1980) and 
fractional (Tsui, St\"ormer, and Gossard 1982) quantum Hall (IQH and FQH) 
effects raised great interest in the properties of a two-dimensional 
electron gas (2DEG) in high magnetic fields.
Both IQH and FQH effects are a manifestation of the occurrence of 
nondegenerate incompressible ground states in the 2DEG spectrum at 
certain (integral for IQH and fractional for FQH) Landau level (LL) 
fillings.
However, unlike the single particle cyclotron gap responsible for the 
IQH effect, the gap separating an FQH incompressible ground state from 
the excited states is due to the electron--electron interactions 
(Laughlin 1983a).
While the occurrence of incompressible ground states of both kinds results
in quantization of the Hall conductance, the origin of incompressible 
ground states in the IQH and FQH effects, i.e.\ the physics underlying the
two quantum Hall effects, is very different.

A simple picture of the FQH states is offered by the mean field composite 
Fermion (CF) approach (Jain 1989, Lopez and Fradkin 1991, Halperin, Lee, 
and Read 1993).
The CF's are obtained in the Chern--Simons (CS) gauge transformation, 
which can be interpreted as attaching to each electron a magnetic flux 
tube oriented opposite to the external magnetic field $B$.
In the mean field approximation, the magnetic field of these flux tubes 
is evenly spread over the occupied area.
If the attached flux tubes carry an even number of flux quanta, the CS
transformation without the mean field approximation leaves the energy 
spectrum and particle statistics unchanged. 
When the mean field approximation is made, the effective magnetic field 
$B^*$ seen by the CF's is lower than the original field $B$ seen by the 
electrons.
The incompressible ground states are predicted to occur at fractional 
electron fillings that correspond to integer fillings of CF LL's.
A gas of strongly interacting electrons is said to behave as a gas of 
weakly interacting CF's, and the FQH effect of electrons is interpreted 
as the IQH effect of CF's.

The mean field CF picture correctly predicts filling factors at which 
the FQH effect has been experimentally observed.
Also, in almost all cases, the mean field CF predictions of low lying 
states of finite systems agree with the results of exact numerical 
calculations in the lowest LL.
However, a very fundamental question: 
`{\em Why does the mean field CF picture work so well?}', 
is not yet completely understood.
The original conjecture that Coulomb and CS gauge interactions beyond 
the mean field cancel each other in the FQH systems cannot possibly be 
correct because the CS interactions are measured on an energy scale 
proportional to $B$, which can be much larger than the energy scale of 
the Coulomb interactions, proportional to $\sqrt{B}$.
Because so many experimental and numerical results in the lowest LL can 
be interpreted in terms of CF's, it is extremely important to understand 
why the CF picture works.

It is known that the CF picture sometimes fails when applied to other 
systems of identical charged Fermions interacting through Coulomb like 
forces.
For example, the occurrence of incompressible states only at some of 
the odd denominator fractional filling factors implies that the CF model 
is not always valid for Laughlin quasiparticles (QP's) in the FQH 
hierarchy (Haldane 1983, Laughlin 1984, Halperin 1984) or for the CF's 
themselves in the CF hierarchy (Sitko {\sl et al.}\ 1996).
The CF picture also fails for electrons in the lowest LL, when the layer
thickness exceeds certain critical value (Shayegan {\sl et al.}\ 1990).
On the other hand, the numerical experiments show that it is correct for 
variety of repulsive interaction potentials (e.g., $V(r)\sim-\ln r$ or 
$r^{-\alpha}$ for $\alpha\ge1$ or even $r^{-1}/\epsilon$ with an 
arbitrary dielectric constant $\epsilon$). 
The original justification of the CF model rested on the assumption that 
spontaneously generated gauge interactions canceled to a substantial
extent the repulsive interactions between electrons, independent of the
exact form of these interactions.
While the CF picture can be used to make certain predictions after it 
has been established that a certain physical system exhibits 
incompressible fluid ground states with Laughlin like correlations 
at appropriate conditions (magnetic field, electron density, layer 
thickness, disorder, material parameters, etc.), it cannot predict 
its own validity for such a system.
Therefore, another very fundamental problem: 
`{\em When does the mean field CF picture work?}', 
needs to be answered.

In this paper, we explain the connection between the form of the Coulomb 
pseudopotential (Haldane 1987) $V(L)$, defined as the dependence of the 
pair interaction energy $V$ on the pair angular momentum $L$, and the 
occurrence of the incompressible ground states in the lowest LL of an
interacting 2DEG at Laughlin--Jain filling factors $\nu={1\over3}$, 
${1\over5}$, ${2\over5}$, ${2\over7}$, etc.
We present arguments justifying the validity of the mean field CF 
picture for the lowest LL and show when and why it can be used.
It is known that the electrons in Laughlin $\nu=(2p+1)^{-1}$ states 
avoid a number ($p$) of pair states with largest repulsion (Haldane 1987).
The origin of incompressible FQH states at certain other filling factors,
such as $\nu={2\over5}$, has been also attributed to the ability of 
avoiding strongly repulsive pair states (Halperin 1983, Haldane 1987, 
Rezayi and MacDonald 1991, Belkhir and Jain 1993).
In order to formally treat the ability to avoid certain pseudopotential 
parameters in the incompressible many body states we use the formalism 
of fractional parentage, well established in the nuclear (Shalit and Talmi
1963) and atomic (Cowan 1981) physics.
It is shown that the condition for the validity of the mean field CF
picture can be more easily expressed in terms of the behavior of the 
pseudopotential $V(L)$ than in terms of the behavior of the 
interaction potential $V(r)$.
The condition on the form of interaction pseudopotential necessary for 
the occurrence of FQH states is given, which defines the class of short 
range pseudopotentials to which mean field CF picture can be applied.
It is shown that the Coulomb interaction in the lowest LL falls in this 
class, while in higher LL's the mean field CF picture can be used only 
below a certain filling factor.
Similarly, the success or failure of the mean field CF picture applied 
to Laughlin QP's, depending on the type of QP's and their filling factor 
(Sitko, Yi, and Quinn 1997), is shown to reflect the behavior of 
appropriate QP pseudopotentials.
It is argued that a QP hierarchy picture taking into account the 
qualitative features of involved pseudopotentials (W\'ojs and Quinn 
2000) should most naturally explain the occurrence and relative 
stability of observed odd denominator FQH states.
We are not discussing even denominator fractions (Willet {\sl et al.}\ 
1987) which are explained in terms of pairing of electrons (Haldane 
and Rezayi 1988, Moore and Read 1991), although a pseudopotential 
approach to the interaction between bound pairs might be possible.
The discussion throughout the paper is illustrated by exact numerical 
calculations of energy spectra and parentage coefficients in Haldane's 
spherical geometry (Haldane 1983), for up to eleven electrons at 
$\nu\sim{1\over3}$ and up to eight electrons at $\nu\sim{1\over5}$ in 
the lowest and excited LL's (matrix dimensions up to $3\cdot10^6$), 
using a modified Lanczos algorithm (Lanczos 1950, Haydock 1980).

The paper is organized as follows.
Section~\ref{secII} gives a brief overview of the numerical (exact 
diagonalization) calculations on the Haldane sphere.
Section~\ref{secIII} explains the mean field CF picture of the FQH states.
The success of the mean field CF approach is illustrated in the energy
spectra of the eight electron system in the lowest LL, for filling 
factors between $\nu={1\over3}$ and ${1\over5}$.
Section~\ref{secIV} introduces the interaction pseudopotential.
Section~\ref{secV} discusses the three electron system.
The idea of fractional parentage from pair states is used to 
characterize the three particle states.
The energy spectra in the lowest and excited LL's are analyzed and 
interpreted in terms of pseudopotential and fractional parentage.
Section~\ref{secVI} generalizes the analysis of the three electron case 
to an arbitrary electron number, and presents the numerical results for 
up to eleven electrons.
Section~\ref{secVII} explains the relation between the form of the 
interaction pseudopotential and the occurrence of many electron 
incompressible ground states.
The Coulomb interaction in different LL's is compared to the harmonic
repulsive interaction and the Coulomb interaction in the atomic shells.
The Hund's rule appropriate for FQH systems is formulated.
The short range pseudopotential is defined, to which the CF model
can be applied.
The prescription for the low energy many electron multiplets is 
derived, which agrees with predictions of the mean field CF picture.
The consequences of the form of pseudopotential for condensation of
QP's in the hierarchy picture is mentioned.
Section~\ref{secVIII} contains the conclusions.

\section{Numerical studies}
\label{secII}

\subsection{Introduction}
\label{secIIa}

In a magnetic field $B$, the lowest LL of a 2DEG can accommodate 
$N_\phi=BC/\phi_0$ electrons per area $C$ ($\phi_0=hc/e$ is the 
magnetic flux quantum).
The measure of electron density is the fraction of occupied states, 
given by the filling factor $\nu=N/N_\phi$, where $N$ is the number
of electrons in the area $C$.
In the absence of electron--electron interactions, the $N_\phi$ single 
particle states are degenerate.
Therefore, these interactions entirely determine the low energy spectrum 
of the system at $\nu<1$ and cannot be treated perturbatively.
Instead, numerical diagonalization techniques have commonly been employed, 
which, however, limit the system to a finite (small) number of electrons.
Different approaches to restrict motion of a finite number of electrons 
to a finite area $C$ to model an infinite 2DEG at a finite density 
include imposing a lateral (parabolic, hard wall, etc.) confinement 
(Laughlin 1983a), using periodic boundary conditions (Haldane and Rezayi 
1985b), or confining electrons on a closed surface (Haldane 1983).
The last approach has proven particularly useful, since it naturally 
avoids edge effects.
Also, the translational symmetry of a (planar) 2DEG is preserved in the 
form of the rotational symmetry of a sphere.
In particular, the pair of good quantum numbers resulting from the 
translational symmetry of a plane: the center of mass and relative 
momenta, correspond to the pair of good quantum numbers on a sphere: 
total angular momentum $L$ and its projection $L_z$ (W\'ojs and Quinn 
1998a).
Consequently, the degeneracies associated with center of mass excitations 
on a plane correspond to those associated with different values of $L_z$ 
($|L_z|\le L$) on a sphere, and the nondegenerate incompressible ground 
states of a planar 2DEG correspond to nondegenerate ($L=0$) ground states 
on a sphere.

\subsection{Haldane sphere}
\label{secIIb}

The magnetic field $B$ perpendicular to the surface of the Haldane 
sphere of radius $R$ is an isotropic radial field produced by 
a magnetic monopole placed at the origin.
The monopole strength $2S$ is defined in the units of elementary flux
$\phi_0=hc/e$, so that the total flux through the sphere is $4\pi BR^2
=2S\phi_0$. 
Dirac's monopole quantization condition requires that $2S$ is an integer
(Dirac 1931), and positive $S$ means magnetic field pointing outwards.
The convenient units of length and energy, magnetic length $\lambda$
and the cyclotron frequency $\hbar\omega_c$, are given by
\begin{eqnarray}
   \lambda^2|S|&=&R^2,
\\
   \hbar\omega_c&=&S{\hbar^2\over\mu R^2}.
\end{eqnarray}
The eigenstates of the single particle Hamiltonian are denoted by 
$\left|S,l,m\right>$ and called monopole harmonics (Wu and Yang 1976).
They are labeled by angular momentum $l$ and its projection $m$.
The degenerate angular momentum shells are equivalent to the LL's
of the planar geometry.
The eigenenergies are given by
\begin{eqnarray}
   E_n&=&{\hbar\omega_c\over2S}\left[l(l+1)-S^2\right] 
\nonumber\\
      &=&\hbar\omega_c\left[n+{1\over2}+{n(n+1)\over2S}\right],
\end{eqnarray}
where the shell (LL) index is defined as $n=l-S=0$, 1, 2, \dots.
The degeneracy of each shell (LL) is $N_\phi=2l+1$.

For the FQH states at filling factors $\nu<1$, only the lowest, spin 
polarized shell (LL) need be considered.
It corresponds to $n=0$ ($l=S$), and for simplicity its single 
particle states will be denoted as $\left|m\right>$.
The spin polarized FQH states in excited LL's will also be studied.
Due to the high (cyclotron) energy of the inter-LL excitations in high
magnetic fields, the FQH states at filling factors $2n<\nu<2n+1$ are 
composed of completely filled LL's (spin up and down) up to the 
$(n-1)$-st one, and a partially filled $n$th LL with the filling 
factor $\nu_n<1$ (we discuss only the spin polarized states in 
partially filled excited LL's).
The Hartree--Fock energy describing interaction between an electron in 
the $n$th LL and the underlying completely filled LL's is a constant.
Therefore, the energy spectrum of $N$ electrons at $\nu_n<1$ in an 
isolated $n$th LL describes (up to this constant) the low energy spectrum 
of $N+2n(2S+n)$ electrons at $\nu=2n+\nu_n$.
Since states of only one LL with a given $n$ appear in the `reduced' 
problem for $\nu=2n+\nu_n$, the following simplified notation will be 
used: filling factor $\nu_n$ will be denoted as $\nu$, and states 
$\left|S,l,m\right>$ will be denoted as $\left|m\right>$.

\subsection{Many body problem}
\label{secIIc}

The object of numerical studies is to diagonalize the electron--electron
interaction Hamiltonian
\begin{equation}
   \hat{H}=\!\!\! \sum_{m_1m_2m_3m_4} \!\!\!\!\!\!
   c^\dagger_{m_1}c^\dagger_{m_2}c_{m_3}c_{m_4}
   \left<m_1,m_2|V|m_3,m_4\right>
\end{equation}
within the Hilbert space ${\cal H}_{\rm MB}$ 
of $N_{\rm MB}=N_\phi![N!(N_\phi-N)!]^{-1}$ degenerate 
antisymmetric $N$ electron states of a given ($N_\phi$-fold 
degenerate) LL.
In the above, $c^\dagger_m$ ($c_m$) creates (annihilates) an electron
in the state $\left|m\right>$.
The two body Coulomb matrix elements have a particularly simple form in 
the lowest LL (Fano {\sl et al.}\ 1986), but they can also be evaluated 
analytically for a general case of inter- or intra-LL scattering.
The $N$ electron Hilbert space ${\cal H}_{\rm MB}$ is spanned by 
single particle configurations $\left|m_1,m_2,\dots,m_N\right>$, 
classified by the total angular momentum projection 
$M=m_1+m_2+\dots+m_N$.
Taking advantage of the Wigner--Eckart theorem, each $(M)$ subspace 
${\cal H}_{\rm MB}(M)$ can be further block diagonalized into $(M,L)$ 
subspaces ${\cal H}_{\rm MB}(M,L)$ corresponding to different values 
of the total angular momentum $L$.
The Wigner--Eckart theorem tells us that because the interaction 
Hamiltonian is a scalar, its matrix element between angular momentum 
eigenstates $\left|L,M,\alpha\right>$ can be written as
\begin{equation}
   \left<L',M',\alpha'|\hat{H}|L,M,\alpha\right>=
   \delta_{LL'}\delta_{MM'}V_{\alpha\alpha'}(L),
\end{equation}
i.e., in terms of a reduced matrix element 
\begin{equation}
   V_{\alpha\alpha'}(L)=\left<L,\alpha'|\hat{H}|L,\alpha\right>
\end{equation}
which is independent of $M$.
Here, index $\alpha$ distinguishes different states in the same 
space ${\cal H}_{\rm MB}(M,L)$.
The typical dimensions are given in table~\ref{tab1}, where we list 
the dimension of the total Hilbert space ${\cal H}_{\rm MB}$, of 
the largest $(M)$ subspace ${\cal H}_{\rm MB}(0)$, of the largest 
$(M,L)$ subspace ${\cal H}_{\rm MB}^{\rm MAX}(M,L)$, and of the 
$(M,L)$ subspace containing the Laughlin $L=0$ ground state, 
${\cal H}_{\rm MB}(0,0)$, for between six and eleven electrons at 
the filling factor $\nu={1\over3}$.
\begin{table}
\caption{
   The dimension $N_\phi$ of the single particle Landau level, 
   dimension $N_{\rm MB}$ of the total many body Hilbert space 
   ${\cal H}_{\rm MB}$,
   dimension $N_{\rm MB}(0)$ of the largest $(M)$ subspace 
   ${\cal H}_{\rm MB}(0)$, 
   dimension $N_{\rm MB}^{\rm MAX}(M,L)$ of the largest $(M,L)$ 
   subspace ${\cal H}_{\rm MB}^{\rm MAX}(M,L)$, 
   and dimension $N_{\rm MB}(0,0)$ of the $(M,L)$ subspace containing 
   the Laughlin $L=0$ ground state, ${\cal H}_{\rm MB}(0,0)$, of $N=6$ 
   to 11 electrons at the filling factor $\nu={1\over3}$.}
\begin{tabular}{rrrrrr}
  $N$ & $N_\phi$ & $N_{\rm MB}$ & $N_{\rm MB}(0)$ & 
  $N_{\rm MB}^{\rm MAX}(M,L)$ & $N_{\rm MB}(0,0)$ \\ \hline
    6 &   16 &         8,008 &       338 &     24 &     6 \\ 
    7 &   19 &        50,388 &     1,656 &     86 &    10 \\ 
    8 &   22 &       319,770 &     8,512 &    352 &    31 \\ 
    9 &   25 &     2,042,975 &    45,207 &  1,533 &    84 \\ 
   10 &   28 &    13,123,110 &   246,448 &  7,069 &   319 \\
   11 &   31 &    84,672,315 & 1,371,535 & 33,787 & 1,160
\end{tabular}
\label{tab1}
\end{table}
Even when both $M$ and $L$ are resolved, exact diagonalization becomes 
difficult when $N$ exceeds 10 and $N_\phi$ exceeds 28.

\begin{figure}[t]
\epsfxsize=3.4in
\epsffile{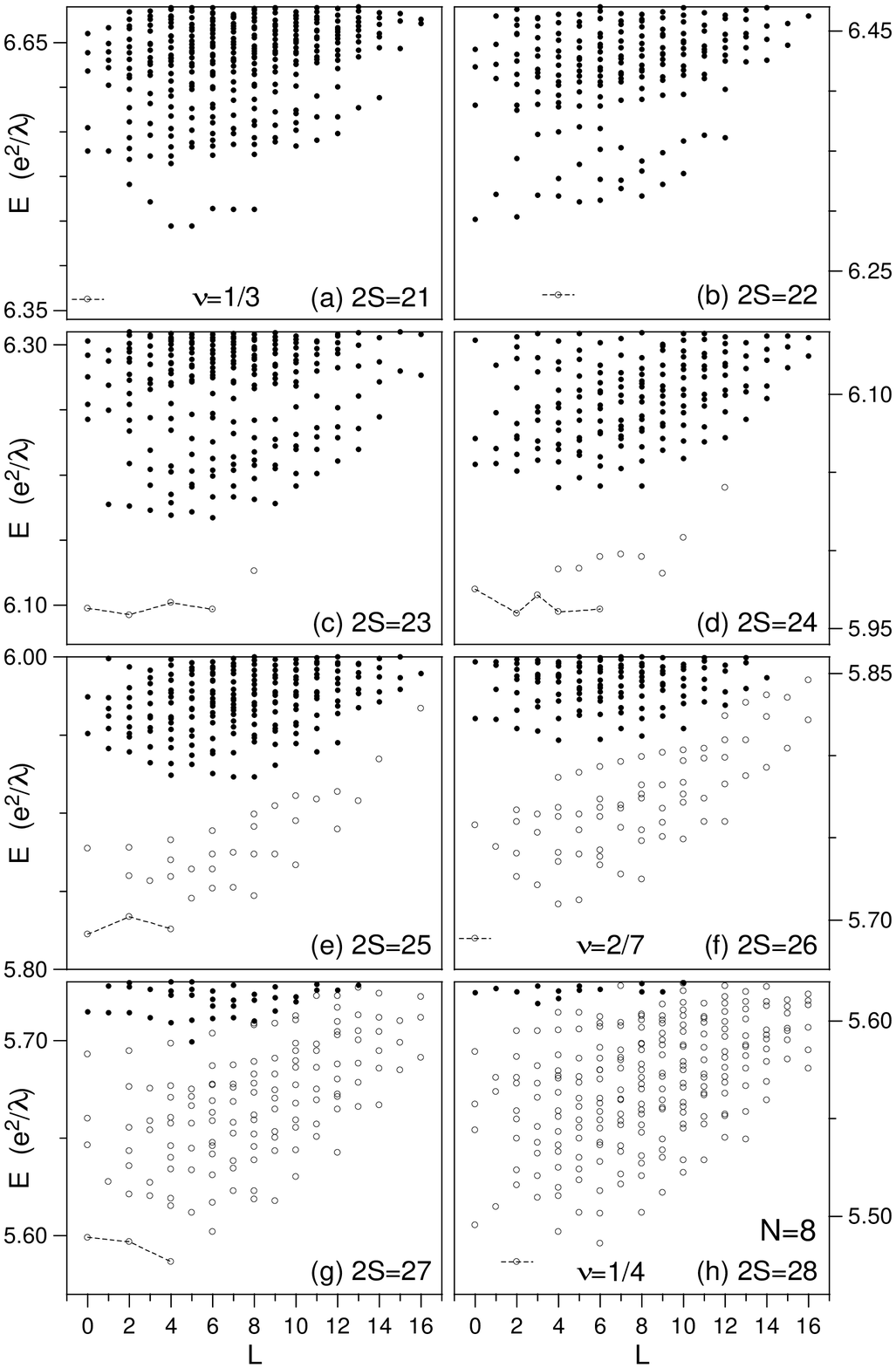}
\caption{
   The energy spectra of eight electrons in the lowest Landau level 
   at the monopole strength $2S$ between 21 and 28. 
   (a) $2S=21$ corresponds to the filling factor $\nu={1\over3}$, 
   the lowest energy state at $L=0$ is the Laughlin ground state; 
   (f) $2S=26$, $\nu={2\over7}$, Jain ground state at $L=0$.
   (h) $2S=28$, $\nu={1\over4}$.
   The low energy states selected by the Chern--Simons transformation 
   with $p=1$ and $p=2$ are marked with open circles and dashed lines, 
   respectively.}
\label{fig1}
\end{figure}

\begin{figure}[t]
\epsfxsize=3.4in
\epsffile{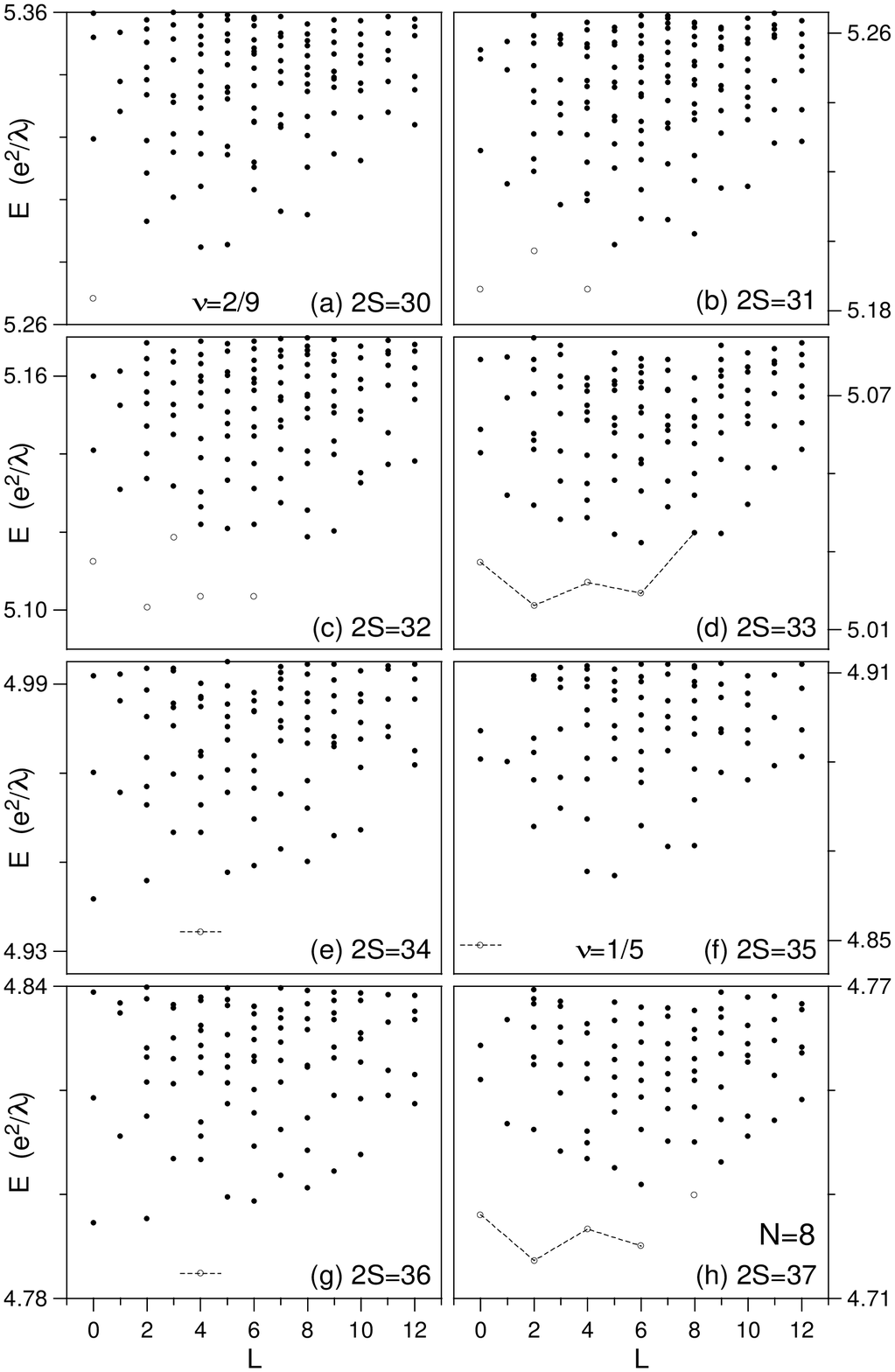}
\caption{
   The energy spectra of eight electrons in the lowest Landau level 
   at the monopole strength $2S$ between 30 and 37. 
   (a) $2S=30$ corresponds to the filling factor $\nu={2\over9}$, 
   the lowest energy state at $L=0$ is the Jain ground state; 
   (f) $2S=35$, $\nu={1\over5}$, Laughlin ground state at $L=0$.
   The low energy states selected by the Chern--Simons transformation 
   with $p=2$ and $p=3$ are marked with open circles and dashed lines, 
   respectively.}
\label{fig2}
\end{figure}

The calculations give the eigenenergies $E$ as a function of total angular 
momentum $L$.
The numerical results for the lowest LL always show one or more $L$ 
multiplets forming a low energy sector (or low energy band).
The spectra for $N$ in the range 6--20 (depending on the filling factor) 
are available in literature and have been extensively analyzed.
As an example, in figures~\ref{fig1} and \ref{fig2} we show the 
energy spectra obtained for eight electrons in the lowest LL, at values 
of $2S$ between 21 and 37; the spectra for $2S<21$ can be found in 
earlier numerical studies (He, Xie, and Zhang 1992).
The Laughlin filling factors $\nu={1\over3}$ and ${1\over5}$ occur 
at $2S=21$ and 35, and the Jain filling factors $\nu={2\over7}$ and 
${2\over9}$ occur at $2S=26$ and 30, respectively.
At $2S=28$, an even denominator filling of $\nu={1\over4}$ occurs.
The low energy bands are marked with open circles.
For some values of $2S$ these bands contain subbands marked with 
dashed lines.
The physical meaning of bands marked in figures~\ref{fig1} and
~\ref{fig2} will be explained in section~\ref{secIIIb}.

\section{Composite Fermion approach}
\label{secIII}

\subsection{Introduction}
\label{secIIIa}

In the Chern--Simons (CS) transformation, an equal and even number 
($2p$) of elementary fluxes $\phi_0$ (a fictitious flux tube of 
strength $2p\phi_0$) oriented opposite to the original magnetic 
field $B$ is attached to each electron.
The composite Fermions (CF's) obtained in this way carry electric 
charge and magnetic flux.
The CS transformation is a gauge transformation and thus the CF energy 
spectrum is identical to the original electron spectrum.

Since attached fluxes are localized on electrons and the magnetic field 
acting on each electron is unchanged, the classical Hamiltonian of the
system is also unchanged.
However, the quantum Hamiltonian includes additional terms describing 
an additional charge--flux (CS) interaction, which arises from the 
Aharonov--Bohm phase attained when one electron's path encircles the 
flux tube attached to another electron.
One difficulty in treatment of the CS interaction results from the fact 
that it contains both two and three body terms; another is the absence
of a small parameter with which to construct a perturbation expansion.

\subsection{Mean field approximation}
\label{secIIIb}

In the mean field approach, the magnetic field due to attached flux 
tubes is evenly spread over the occupied area $C$.
The mean field CF's obtained in this way move in an effective magnetic 
field $B^*=B-2p\phi_0\,N/C$.
An effective filling factor $\nu^*$ seen by one CF is defined as 
\begin{equation}
   (\nu^*)^{-1}=\nu^{-1}-2p,
\label{eqjain}
\end{equation}
so that
\begin{equation}
   B^*\nu^*=B\nu={N\over C}\phi_0.
\end{equation}
Negative $\nu^*$ means negative $B^*$ (oriented opposite to $B$).
It has been shown that the mean field Hamiltonian of noninteracting 
CF's gives a good qualitative description of the low lying states of 
interacting electrons in the lowest LL.
The Jain sequence of incompressible ground states is predicted at 
filling factors $\nu$ for which $\nu^*$ is an integer (Jain 1989), 
and the $\nu^*=1$ states correspond to Laughlin $\nu=(2p+1)^{-1}$ 
states (Laughlin 1983a).
If $\nu^*$ is not an integer, the low lying states contain a number 
of QP's ($N_{\rm QP}\le N$) in the neighboring incompressible state 
with integer $\nu^*$.

On a sphere, an effective CF monopole strength is 
\begin{equation}
   2S^*=2S-2p(N-1),
\label{eqcfms}
\end{equation}
and $l^*=|S^*|$ plays the role of the angular momentum of the lowest 
CF shell (Chen and Quinn 1996).
If $n$ lowest CF LL's at $2S^*$ are filled completely by $N$ CF's, 
the corresponding $N$ electron state at $2S$ is incompressible.
The states at other values of $2S$ are compressible and contain 
$N_{\rm QP}$ QP's in the neighboring incompressible state of an 
equal number of $N$ electrons at $2S_{\rm INC}$,
\begin{equation}
   N_{\rm QP}=n(|2S_{\rm INC}^*|-|2S^*|).
\label{eqnqps}
\end{equation}
Here $2S_{\rm INC}^*$ is the effective monopole strength calculated for 
the incompressible state, i.e.\ $2S_{\rm INC}^*=2S_{\rm INC}-2p_{\rm INC}
(N-1)$, and $n$ is an integral number of completely filled CF LL's.
Positive $N_{\rm QP}$ corresponds to quasielectrons (QE's) in the 
$(n+1)$-st (lowest unoccupied) CF shell, each with angular momentum 
$l_{\rm QE}=l^*+n$.
Negative $N_{\rm QP}$ corresponds to quasiholes (QH's) in the $n$th 
(highest occupied) CF shell, each with angular momentum $l_{\rm QH}=
l^*+n-1$.
Different values of $2S$ that lead to the same value of $l^*=|S^*|$ are 
equivalent and their low energy bands contain the same $L$ multiplets.

It is noteworthy that the CS transformation applied to the state at $2S$ 
can have a different flux strength ($2p$) than that ($2p_{\rm INC}$) 
applied to the incompressible state $2S_{\rm INC}$.
Consequently, alternative pictures of the $(N,2S)$ state, containing 
different numbers and/or types of QP's, can be obtained (Yi {\sl et al.}\ 
1996).
Writing $p_{\rm INC}$ and $p$ explicitly, equation~\ref{eqnqps} can 
be written as 
\begin{equation}
   N_{\rm QP}=n\left(
      \left|2S_{\rm INC}-2p_{\rm INC}(N-1)\right|-
      \left|2S-2p(N-1)\right|
   \right).
\label{eqnqps2}
\end{equation}
The original spectrum of interacting electrons is similar to that of 
noninteracting mean field CF's in a sense that (i) the lowest band of 
angular momentum multiplets contains states of the minimum number of 
QP's consistent with the values of $N$ and $2S$, and (ii) neighboring 
excited bands contain additional QE--QH pairs.

Let us illustrate the success of the mean field CF approach in predicting 
the lowest band of multiplets on the example of an eight electron system.
The sequence of incompressible states is given in table~\ref{tab2}.
\begin{table}
\caption{
   The incompressible states of eight electrons; 
   filling factor $\nu\ge{1\over5}$.}
\begin{tabular}{r|cc|cc}
       & \multicolumn{2}{c|}{$n=1$, $2S^*=7$}
       & \multicolumn{2}{c }{$n=2$, $2S^*=2$} \\ \hline
  $ p$ & $|2S|$ & $\nu$ & $|2S|$ & $\nu$ \\ \hline
  $-2$ &   21   &  1/3  &   26   &  2/7  \\ 
  $-1$ &    7   &   1   &   12   &  2/3  \\ 
  $ 0$ &    7   &   1   &    2   &   2   \\ 
  $ 1$ &   21   &  1/3  &   16   &  2/5  \\ 
  $ 2$ &   35   &  1/5  &   30   &  2/9 
\end{tabular}
\label{tab2}
\end{table}
Eight mean field CF's fill completely one CF LL ($n=1$) at $|2S^*|=7$ 
and two CF LL's ($n=2$) at $|2S^*|=2$.
Following equation~\ref{eqcfms}, the sequences of incompressible states 
for CF fillings $n=1$ and 2 are generated by varying $p=0$, $\pm1$, 
$\pm2$, \dots.
States listed in table~\ref{tab2} ($\nu=2$, 1, ${2\over3}$, ${2\over5}$, 
${1\over3}$, ${2\over7}$, ${2\over9}$, and ${1\over5}$) are all the 
incompressible eight electron states at filling factors greater than 
or equal to $\nu={1\over5}$ (filling of more than two CF LL's requires 
$N>8$).
The states outside the incompressible sequence of $2S_{\rm INC}=2$, 
7, 12, 16, 21, 26, 30, 35, \dots\ are compressible and contain an 
appropriate number of QP's, given by equation~\ref{eqnqps}.

The spectra of an eight electron system in the lowest LL for values 
of $2S$ between 21 and 37, i.e.\ for the filling factors $\nu$ from 
${1\over3}$ down to below ${1\over5}$, are shown in figures~\ref{fig1} 
and \ref{fig2}.
In figure~\ref{fig1}, the open circles and dashed lines mark bands 
of multiplets predicted in the mean field CF picture as the lowest 
energy states of CF's for $p=1$ and $p=2$, respectively.
In figure~\ref{fig2}, all shown states belong to the lowest band 
corresponding to $p=1$, and the open circles and dashed lines mark 
bands obtained for $p=2$ and $p=3$, respectively.
The range of $2S$ shown in figure~\ref{fig1} alone covers all values 
of $l^*$ from $N-1$ to 0 (for $p=1$) and thus exhausts all possible 
configurations of QP's for the eight electron system.
Let us analyze the spectra in figure~\ref{fig1} in greater detail.

At $2S=21$ the CS transformation with $p=1$ gives $2S^*=7$.
The lowest CF LL is completely filled ($\nu^*=n=1$) and the Laughlin 
incompressible $\nu_{\rm INC}={1\over3}$ state with $L=0$ is formed.
The CS transformation with $p=2$ gives $2S^*=-7$ and the equivalent 
interpretation of the ground state.
At $2S=22$ the CS transformation with $p=1$ gives $2S^*=8$.
The lowest CF LL has degeneracy of $2S^*+1=9$ so it holds $N=8$ CF's 
and one QH with $l_{\rm QH}=4$ (QH in the $\nu_{\rm INC}={1\over3}$ 
state).
Therefore, the low energy band contains a single multiplet with $L=4$.
The CS transformation with $p=2$ gives $2S^*=-6$ which corresponds to
a completely filled lowest CF LL and one QE with $l_{\rm QE}=4$ in the 
first excited CF LL.
Depending on the applied CS transformation, the $L=4$ ground state can 
be viewed as a state of either a single QE or a single QH in the 
appropriate CF LL (Yi {\sl et al.}\ 1996).
The low energy multiplets obtained using the CS transformation with 
$p=1$ at $2S=23$, 24, \dots, 28 contain 2, 3, \dots, 7 QH's in the 
lowest CF LL (i.e.\ in the $\nu_{\rm INC}={1\over3}$ state), each 
with angular momentum $l_{\rm QH}={9\over2}$, 5, \dots, 7, respectively.
For example, at $2S=24$ the band of states of three QH' each with
$l_{\rm QH}=5$ contains the following multiplets: $L=0$, 2, 3, $4^2$, 
5, $6^2$, 7, 8, 9, 10, and 12.
At $2S\ge23$ the CS transformation with $p=2$ selects a subset of
multiplets out of those obtained with $p=1$, and the low energy subband 
corresponding to $p=2$ develops in the $p=1$ band.
For example, the low energy $p=2$ subband predicted for $2S=23$ 
($2S^*=-5$) contains two QE's each with $l_{\rm QE}={7\over2}$, and 
thus $L=0$, 2, 4, and 6.
At $2S=26$ the CF monopole strength for $p=2$ is $2S^*=-2$ and 
two lowest CF LL's are completely filled ($\nu^*=n=2$).
The ground state is the incompressible Jain $\nu_{\rm INC}={2\over7}$ 
state with $L=0$.
At $2S=25$ the CF monopole strength for $p=2$ is $2S^*=-3$ and at 
$2S=27$ it is $2S^*=-1$.
In both cases, the low energy band contains two QP's each with 
$l_{\rm QP}={5\over2}$ in the $\nu_{\rm INC}={2\over7}$ state (two QE's 
at $2S=25$ and two QH's at $2S=27$).
For $2S=24$ one obtains $2S^*=-4$ and the lowest energy band contains 
three QE's each with $l_{\rm QE}=3$.
Finally, for $2S=28$ one obtains $2S^*=0$ and one QH with $l_{\rm QH}=2$ 
in the second excited CF LL. 
The effective magnetic field acting on the CF's vanishes, and this state 
is assigned an even denominator filling factor $\nu={1\over4}$.

Higher energy bands, containing multiplets with additional QE--QH pairs,
are more difficult to identify in figures~\ref{fig1} and \ref{fig2} 
than the lowest ones.
However, for $2S=21$ one can easily notice the low lying band of states 
at $L=2$, 3, 4, 5, 6, 7, and 8, which correspond to the states of one 
QE--QH pair ($l_{\rm QE}={9\over2}$ and $l_{\rm QH}={7\over2}$) in the 
mean field CF picture.
Similarly, the band of QE--QH pair states for $2S=26$ occurs at $L=2$, 
3, 4, and 5 ($l_{\rm QE}=3$ and $l_{\rm QH}=2$).
For $2S=25$ the lowest band contains two QH's each with $l_{\rm QH}
={5\over2}$ in the $\nu^*=2$ CF state ($L=0$, 2, and 4).
The first excited band has two subbands at the same CF energy.
One contains states corresponding to three QH's each with $l_{\rm QH}=
{5\over2}$ and one QE with $l_{\rm QE}={7\over2}$.
The allowed multiplets of such QP system are $L=1^2$, $2^3$, $3^3$, $4^3$,
$5^3$, $6^2$, 7, and 8.
The other contains states of one QH in the lowest CF LL ($l_{\rm QH1}=
{3\over2}$) and one QH in the first excited CF LL ($l_{\rm QH2}={5\over2}$).
The allowed multiplets in this subband are $L=1$, 2, 3, and 4.
One can identify in figure~\ref{fig1}(f) a few multiplets with highest
angular momenta ($L=8$, 7, $6^2$, \dots) of this band.

\begin{figure}[t]
\epsfxsize=3.4in
\epsffile{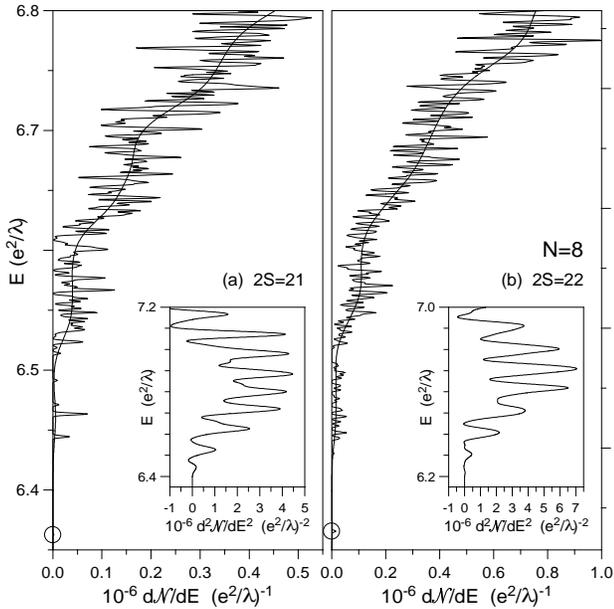}
\caption{
   The density of states $d{\cal N}/dE$ for the eight electron 
   spectra at $2S=21$ (a) and 22 (b).
   The thin and thick lines correspond to two different broadenings 
   of discrete energy levels.
   Inset: the differential density of states $d^2{\cal N}/dE^2$.
   The plateaus in $d{\cal N}/dE$ and the minima in $d^2{\cal N}/dE^2$
   correspond to the bands of states with an increasing number of 
   quasielectron--quasihole pairs.}
\label{fig3}
\end{figure}

The bands of states containing an increasing number of QE--QH pairs are 
much better visible in the density of states (DOS), $d{\cal N}(E)/dE$,
plotted in figure~\ref{fig3}.
Frames (a) and (b) show the data for $2S=21$ (Laughlin $\nu={1\over3}$ 
ground state) and $2S=22$ (one QH in the ground state), respectively.
The continuous DOS is obtained by broadening of discrete energy levels 
with Gaussians,
\begin{equation}
   {d{\cal N}(E)\over dE}=
   {\sqrt{\pi}\over\delta}
   \sum_{L\alpha}(2L+1)\exp-{|E-E_{L\alpha}|^2\over\delta^2},
\end{equation}
where summation goes over all $L$ multiplets (distinguished by 
different $\alpha$), and the normalization prefactor guarantees that 
$\int [d{\cal N}(E)/dE]\,dE={\cal N}$, the total number of states.
The thin lines were obtained for $\delta=0.001\,e^2/\lambda$ and the 
thick lines correspond to $\delta=0.02\,e^2/\lambda$.
The thick lines, free of noise characteristic of the discrete spectrum,
reveal a series of equidistant peaks and/or steps in the DOS.
The peaks corresponding to the ground states are hardly visible and 
their positions have been marked with open circles.
A number of higher peaks (at lower energies) or plateaus plateaus 
(at higher energies) are the remnants of the CF bands with increasing 
numbers of QP's.
The quasiperiodic character of the DOS spectrum is even more pronounced 
in the derivatives of the DOS, shown in the insets (calculated only for 
$\delta=0.02\,e^2/\lambda$).
The plateaus in $d{\cal N}/dE$ correspond to the minima in $d^2{\cal 
N}/dE^2$, and the average distance between the neighboring ones is 
about $0.094\,e^2/\lambda$.
In the mean field CF picture, this quantity is interpreted as the energy 
of an QE--QH pair in the Laughlin $\nu={1\over3}$ ground state.

The Fermi liquid picture can be further applied to the QP's
(Sitko {\sl et al.}\ 1996).
The incompressible state is treated as a `vacuum' state, and the QP's
created in this state interact with one another through appropriate
pseudopotentials.
The pseudopotentials were determined by studying the energy spectra 
corresponding to two QP's, and then used to calculate the QP--QP 
interaction energy in states corresponding to a larger number of QP's.
Good agreement with the actual low energy bands of the electron systems 
was obtained.

\subsection{Energy scales and fluctuations 
            beyond mean field approximation}
\label{secIIIc}

\begin{figure}[t]
\epsfxsize=3.4in
\epsffile{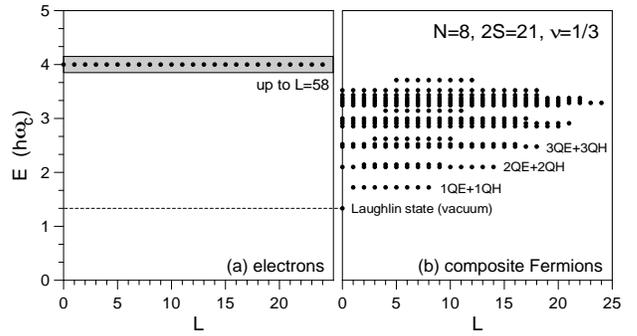}
\caption{
   The energy spectra of noninteracting electrons (a) and noninteracting 
   composite Fermions (b). 
   The characteristic energy of the Coulomb interaction is marked
   in frame (a) with a shaded rectangle.}
\label{fig4}
\end{figure}

Despite the success of the mean field CF approach in describing the low 
energy spectra of interacting electrons in many numerical (exact) 
calculations carried out for finite systems, the reason for its success 
still remains a puzzle.
The original conjecture that the CF transformation converts a system of 
strongly interacting electrons into one of weakly interacting CF's
cannot possibly be correct because the CS interactions among fluctuations 
are measured on an energy scale proportional to $\hbar\omega_c\propto B$,
which can be much larger than the energy scale of the Coulomb interactions,
proportional to $e^2/\lambda\propto\sqrt{B}$.
This is demonstrated in figure~\ref{fig4}, where the original energy 
spectrum of free electrons is compared to that of noninteracting mean 
field CF's (note that the degeneracy of multiplets is not shown).
Clearly, inclusion of the electron--electron Coulomb interaction with 
characteristic energy as small as marked in figure~\ref{fig4} with 
a shaded rectangle cannot reproduce the separation of levels present 
in the mean field CF spectrum.
Because so many results can be successfully interpreted in terms of 
composite Fermions, the understanding of the actual reason for the
success of the mean field CF model, as well as defining its limitations 
and range of applicability, is extremely important.

\section{Pseudopotential of Coulomb interaction}
\label{secIV}

The two body interaction Hamiltonian of the many body system can be 
expressed as 
\begin{equation}
   \hat{H}=\sum_{i<j} \sum_{L} V(L)\,\hat{\cal P}_{ij}(L).
\label{hproj}
\end{equation}
Here, $V(L)$ is the two particle interaction pseudopotential (Haldane 
1987) defined as the interaction energy of a pair in the eigenstate 
$\left|L\right>$ of angular momentum $L$,
\begin{equation}
   \hat{H}\left|L\right>=V(L)\left|L\right>, 
\end{equation}
and $\hat{\cal P}_{ij}(L)$ is the projection operator onto the subspace 
with the pair $ij$ in the state $\left|L\right>$.
Pair angular momentum $L$ measures the average squared electron--electron 
distance $d^2$.
It can be shown that within the $n$th LL of the Haldane sphere
\begin{equation}
   {\hat{d}^2\over R^2}=
   2+{S^2\over l(l+1)}\left(2-{\hat{L}^2\over l(l+1)}\right).
\label{eqharm}
\end{equation}
Notice that $0<d^2<(2R)^2$ and $d^2\equiv2R^2$ for $2S=0$.

\begin{figure}[t]
\epsfxsize=3.4in
\epsffile{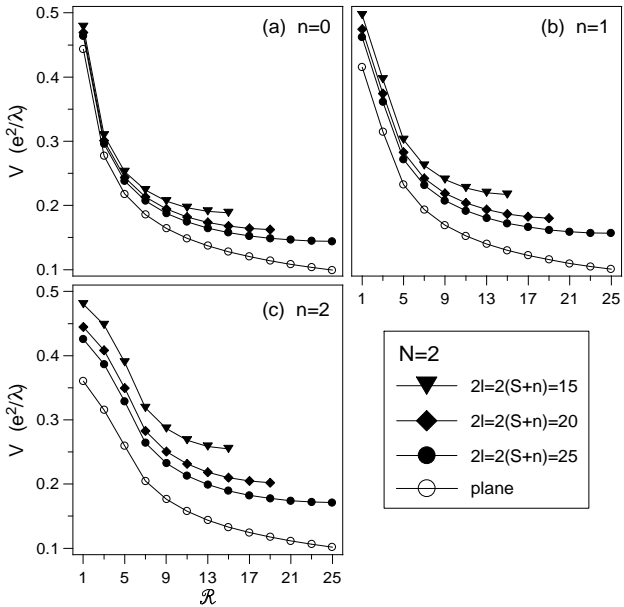}
\caption{
   The pseudopotentials $V$ of the Coulomb interaction in the lowest 
   (a), first excited (b), and second excited (c) Landau levels, 
   as a function of the relative angular momentum ${\cal R}$. 
   Open circles: plane; full triangles, diamonds and circles: 
   Haldane sphere with $l={15\over2}$, 10, and ${25\over2}$, respectively.}
\label{fig5}
\end{figure}

Due to the confinement of single particle states to one (lowest) LL, 
the number of pair states is strongly limited, and the electron--electron 
interaction potential enters the Hamiltonian $H$ only through a small 
set of pseudopotential parameters.
This reveals the magnetic field quantization of electron--electron 
interaction, i.e., electron--electron separation (Laughlin 1983b).
On a Haldane sphere with a given $2S$, a finite number of these 
parameters, $V(2l-{\cal R})$, where ${\cal R}\le2l$ is an odd integer, 
determines many body eigenstates and eigenenergies.
Using the relative angular momentum ${\cal R}$ instead of the
eigenvalue $L$ of total angular momentum $\hat{\bf L}=\hat{\bf l}_1+
\hat{\bf l}_2$ to label pair states and pseudopotential coefficients 
allows for meaningful comparison of the pseudopotentials in the planar 
system and in spherical systems with different $l$ (or $2S$).
On a sphere, ${\cal R}$ is defined as 
\begin{equation}
   {\cal R}=2l-L,
\end{equation}
and on a plane it is equal to the angular momentum associated with 
the relative coordinate ${\bf r}={\bf r}_1-{\bf r}_2$.
In both cases, larger ${\cal R}$ means larger separation (see 
equation~\ref{eqharm} for the sphere).
Figure~\ref{fig5} shows pseudopotentials $V({\cal R})$ calculated 
for the lowest and the first two excited LL's ($n=0$, 1, and 2) for 
the plane and for the Haldane sphere with $l={15\over2}$, 10, and 
${25\over2}$.
All pseudopotentials $V({\cal R})$ in figure~\ref{fig5} decrease 
with increasing ${\cal R}$.

The important part of the pseudopotential spectrum is where its slope 
is the highest.
It follows from equation~\ref{eqharm} that each pair state with a given 
$L$ corresponds to a certain average separation $d$ and, roughly, 
$d\propto{\cal R}$.
Large slope $dV/d{\cal R}$ means large energy gradient, i.e.\ large 
effective force, that would describe two point charges at a distance $d$.
This effective force is solely due to the Coulomb force, but takes into 
account different spread of electron wavefunctions in pair states for 
different $2S$, $l$, and ${\cal R}$.
As will be shown later, the crucial difference between the lowest LL 
(a) and excited LL's (b,c) is that in the former case $V({\cal R})$ 
decreases more quickly at the smallest values of ${\cal R}$.

Let us define a model hard core pseudopotential $V_{\rm HC}$ for which
\begin{eqnarray}
   V_{\rm HC}({\cal R}  ) &\gg& 
   V_{\rm HC}({\cal R}+2), 
\nonumber\\
   V_{\rm HC}({\cal R}-2) - V_{\rm HC}({\cal R}  ) &\gg& 
   V_{\rm HC}({\cal R}  ) - V_{\rm HC}({\cal R}+2)
\label{eqvhc}
\end{eqnarray}
for all values of ${\cal R}$.
The $V_{\rm HC}$ is an `ideal' short range pseudopotential (the class 
of short range pseudopotentials leading to the similar, Laughlin like 
short range correlations will be formally defined in section~\ref{secVIIe}).
The conditions~\ref{eqvhc} can be rewritten as $dV/d{\cal R}\ll0$ and 
$d^2V/d{\cal R}^2\gg0$, where the derivatives are to be understood as 
finite differences.
Clearly, in the low lying many body eigenstates of $V_{\rm HC}$, 
electrons must avoid as much as possible pair states with largest 
repulsion, i.e.\ pair states with the smallest separation or smallest 
values of ${\cal R}$.
The many body states which avoid certain values of ${\cal R}$ can be 
constructed explicitly using parentage or grandparentage coefficients.
In the following sections we shall investigate in detail the connection 
between the low lying states of the FQH systems and the avoiding of 
pair states with largest repulsion.

\section{Three electron system}
\label{secV}

\subsection{Coefficients of fractional parentage}
\label{secVa}

We begin the discussion of the three electron case by listing in 
table~\ref{tab3} all possible $L$ multiplets appearing in the spectrum 
for a given single particle angular momentum $l$.
\begin{table}
\caption{
   The number of times an $L$ multiplet appears for a system of 
   three electrons of angular momentum $l$.
   Top: even values of $2l$; bottom: odd values of $2l$.
   Blank spaces are equivalent to zeros.}
\begin{tabular}{r|ccccccccccccccccccc}
  $_{2l}\mbox{}^{2L}$
     &0&2&4&6&8&10&12&14&16&18&20&22&24&26&28&30&32&34&36\\\hline
    2&1&&&&&&&&&&&&&&&&&&\\
    4&&1&&1&&&&&&&&&&&&&&&\\
    6&\underline{1}&&1&1&1&&1&&&&&&&&&&&&\\
    8&&\underline{1}&&2&1&1&1&1&&1&&&&&&&&&\\
   10&\underline{\underline{1}}&&\underline{1}&\underline{1}&2&1&2&1
      &1&1&1&&1&&&&&&\\
   12&&\underline{\underline{1}}&&\underline{2}&\underline{1}&2&2&2
      &1&2&1&1&1&1&&1&&&\\
   14&\underline{\underline{\underline{1}}}&&\underline{\underline{1}}
      &\underline{\underline{1}}&\underline{2}&\underline{1}
      &3&2&2&2&2&1&2&1&1&1&1&&1
\end{tabular}
\begin{tabular}{r|ccccccccccccccccccc}
  $_{2l}\mbox{}^{2L}$
     &1&3&5&7&9&11&13&15&17&19&21&23&25&27&29&31&33&35&37\\\hline
    3&&1&&&&&&&&&&&&&&&&&\\
    5&&1&1&&1&&&&&&&&&&&&&&\\
    7&&\underline{1}&1&1&1&1&&1&&&&&&&&&&&\\
    9&&\underline{1}&\underline{1}&1&2&1&1&1&1&&1&&&&&&&&\\
   11&&\underline{\underline{1}}&\underline{1}&\underline{1}&2&2&1&2&1
      &1&1&1&&1&&&&&\\
   13&&\underline{\underline{1}}&\underline{\underline{1}}&\underline{1}
      &\underline{2}&2&2&2&2&1&2&1&1&1&1&&1&&
\end{tabular}
\label{tab3}
\end{table}
An eigenfunction of three electrons each of angular momentum $l$ 
whose total angular momentum is $L$ will be denoted by 
$\left|l^3,L\alpha\right>$, with an index $\alpha$ distinguishing 
different multiplets with the same $L$.
This state can be written as 
\begin{equation}
   \left|l^3,L\alpha\right>=\sum_{L_{12}}F_{L\alpha}(L_{12})
   \left|l^2,L_{12};l,L\right>,
\label{eq1}
\end{equation}
a combination of product states $\left|l^2,L_{12};l,L\right>$ in 
which $l_1=l_2=l$ are added to obtain pair angular momentum $L_{12}$, 
and then $l_3=l$ is added to $L_{12}$ to obtain total angular 
momentum $L$ (Shalit and Talmi 1963, Cowan 1981).
Note that state $\left|l^3,L\alpha\right>$ is antisymmetric under 
interchange of any pair of particles 1, 2, and 3, while states 
$\left|l^2,L_{12};l,L\right>$ are antisymmetric only under interchange 
of particles 1 and 2.
The factor $F_{L\alpha}(L_{12})$, or $F_{L\alpha}({\cal R})$ where 
${\cal R}=2l-L_{12}$, is called the coefficient of fractional parentage 
(CFP) associated with pair angular momentum $L_{12}$. 

The two particle interaction matrix element can be conveniently 
expressed through the CFP's and the pseudopotential coefficients 
(Sitko {\sl et al.}\ 1996),
\begin{equation}
   \left<l^3,L\alpha\right|V\left|l^3,L\beta\right>
   =3\sum_{\cal R} F_{L\alpha}({\cal R})F_{L\beta}({\cal R})\,V({\cal R}).
\label{eq2}
\end{equation}
If state $\left|l^3,L\alpha\right>$ is an eigenstate of the interacting
system, its energy is
\begin{equation}
   E_{L\alpha}=3\sum_{\cal R}{\cal F}_{L\alpha}({\cal R})\,V({\cal R}),
\label{eq2a}
\end{equation}
where ${\cal F}_{L\alpha}=|F_{L\alpha}|^2$.
The CFP's for three particles with given $l$ can be derived analytically
or found in nuclear (Shalit and Talmi 1963) or atomic (Cowan 1981)
physics books.
Note however that the squared CFP's, which appear in equation~\ref{eq2a} 
and measure the probability that a pair of electrons $ij$ are in the 
pair state of angular momentum ${\cal R}$ can be expressed as
\begin{equation}
   {\cal F}_{L\alpha}({\cal R})=
   \left<L\alpha\right|\hat{\cal P}_{ij}({\cal R})\left|L\alpha\right>.
\label{eqcalFa}
\end{equation}
It follows from equation~\ref{hproj} that they can be calculated quite 
easily for any state $\left|L\alpha\right>$ as the expectation value 
of the `selective interaction' Hamiltonian $\hat{H}_{\cal R}$, whose 
only nonvanishing pseudopotential parameter is $V({\cal R})=1$,
\begin{equation}
   {\cal F}_{L\alpha}({\cal R})={1\over3}
   \left<L\alpha\right|\hat{H}_{\cal R}\left|L\alpha\right>.
\label{eqcalFb}
\end{equation}

\subsection{Hard core repulsive interaction}
\label{secVb}

For the hard core pseudopotential defined in equation~\ref{eqvhc}, 
the low lying states must avoid low values of ${\cal R}$ as much as 
possible within the available Hilbert space.
They have the maximum allowed number of vanishing CFP's which 
correspond to lowest values of ${\cal R}$, ${\cal F}_{L\alpha}(1)=
{\cal F}_{L\alpha}(3)=\dots=0$.
In such states, all pairs $ij$ have zero projection onto pair states 
with a number of lowest values of ${\cal R}$,
\begin{equation}
   \sum_{i<j} \sum_{{\cal R}=1,3,\dots} \hat{\cal P}_{ij}({\cal R})
   \left|L\alpha\right>=0,
\end{equation}
or with a number of pseudopotential parameters associated with the 
strongest repulsion, $V(1)$, $V(3)$, \dots.

For three electrons (Fermions), the angular momenta of states in which 
${\cal R}\ge3$, 5, \dots, for all pairs can be predicted from the 
following argument (W\'ojs and Quinn 1998b).
If we choose ${\cal R}=1$ for the pair of electrons 1 and 2 (i.e.\ 
$L_{12}=2l-1$), and add to $L_{12}$ the same single particle angular 
momentum $l$ of the third electron, then the total angular momentum 
$L$ must satisfy the vector addition rule, $|L_{12}-l|\le L\le L_{12}+l$.
The antisymmetrization of the total wavefunction will eliminate some 
of the values of $L$ from this range, but it is guaranteed that no
states with $L$ smaller than the minimum value, $L<l-1$, can have 
nonvanishing parentage from ${\cal R}=1$.
In table~\ref{tab3}, we have underlined the three electron states with 
$L<l-1$, which must therefore have ${\cal R}\ge3$ for all pairs.
The next higher value of ${\cal R}$ to avoid is 3, and, using the same 
argument as above, we obtain that all states with $L<l-3$ must have 
${\cal R}\ge5$ (double underlined in table~\ref{tab3}).
Further, states with $L<l-5$ must all have ${\cal R}\ge7$ (triple 
underlined in table~\ref{tab3}), and so on.
In table~\ref{tab4} we list the values of $2L$ for which the CFP with 
${\cal R}=1$ or with ${\cal R}\le3$ or with ${\cal R}\le5$ must vanish,
i.e. ${\cal R}\ge3$, 5, or 7, respectively.
\begin{table}
\caption{
   The allowed values of $2L$ for a three electron system that must 
   have ${\cal R}\ge3$, 5, and 7.
   The listed values correspond to the underlined $L$ multiplets 
   in table~\protect\ref{tab3}.}
\begin{tabular}{c|lllllllll}
  $2l$   & 6& 7& 8& 9&10&11&12&13&14 \\ \hline
  $2L\,({\cal R}\ge3)$ & 0& 3& 2& 3,5& 0,4,6& 3,5,7& 2,6$^2$,8
            & 3,5,7,9$^2$& 0,4,6,8$^2$,10 \\
  $2L\,({\cal R}\ge5)$ & & & & & 0& 3& 2& 3,5& 0,4,6 \\
  $2L\,({\cal R}\ge7)$ & & & & & & & & & 0
\end{tabular}
\label{tab4}
\end{table}
The $L=0$ states for $2S=6$, 10, and 14 are the Laughlin ground states 
with $\nu={1\over3}$, ${1\over5}$, and ${1\over7}$, respectively.

Note that the multiplets listed at $2l$ with ${\cal R}\ge{\cal R}_{\rm 
MIN}$ are always the same as those at $2l-2p(N-1)$ with ${\cal R}\ge{\cal 
R}^{\rm MIN}-2p$.
But for the lowest LL ($l=S$), $2S-2p(N-1)$ is just $2S^*$, the effective 
monopole strength of CF's!
This very important result remains true for any number of electrons, 
and will be discussed in more detail in section~\ref{secVIb}.

At $2S=8$, two $L=3$ multiplets occur (see table~\ref{tab3}) and the 
interparticle interaction must be diagonalized in this two dimensional 
subspace.
The CFP for ${\cal R}=1$ does not vanish identically in entire subspace 
because $L\ge l-1$.
However, a linear combination can be constructed for which it does.
For a model pseudopotential with $V(1)>0$ and all other parameters 
vanishing, this would be the lower (zero energy) eigenstate.
At $2S=14$ there are three allowed $L=6$ multiplets, out of which one 
linear combination can be constructed with zero CFP for both ${\cal R}
=1$ and 3, and another one without CFP for ${\cal R}=1$ but with 
significant CFP for ${\cal R}=3$.

\subsection{Coulomb interaction in lowest and excited Landau levels}
\label{secVc}

\begin{figure}[t]
\epsfxsize=3.4in
\epsffile{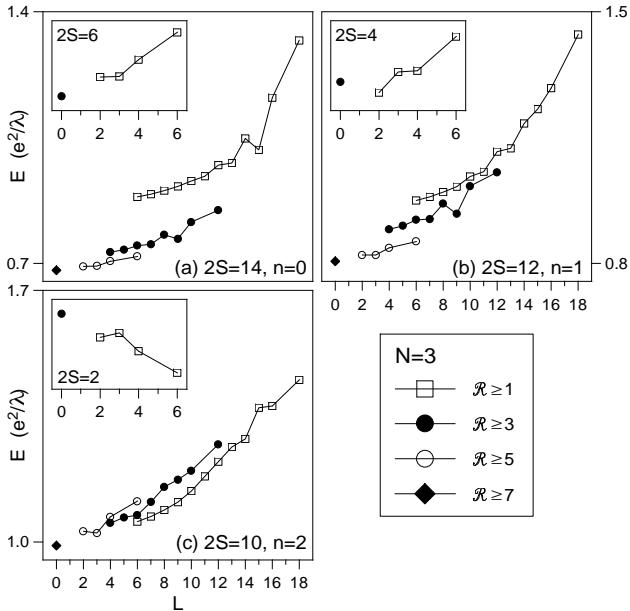}
\caption{
   The Coulomb energy of three electrons each with $l=7$ in the 
   lowest (a), first excited (b), and second excited (c) Landau level.
   Diamonds: states with ${\cal R}\ge7$, 
   i.e.\ ${\cal F}(1)\approx{\cal F}(3)\approx{\cal F}(5)\approx0$ and 
   ${\cal F}(7)>0$;
   open circles: ${\cal R}\ge5$, 
   i.e.\ ${\cal F}(1)\approx{\cal F}(3)\approx0$ and ${\cal F}(5)>0$;
   full circles: ${\cal R}\ge3$, 
   i.e.\ ${\cal F}(1)\approx0$ and ${\cal F}(3)>0$; and
   squares: ${\cal R}\ge1$, 
   i.e.\ ${\cal F}(1)>0$.
   The ground states in all frames are the Laughlin $\nu={1\over7}$ 
   states within different LL's.
   Insets: spectra for $l=3$; 
   the ground state for $n=0$ is the Laughlin $\nu={1\over3}$ state.}
\label{fig6}
\end{figure}

How does this work for the actual Coulomb interaction?
Figure~\ref{fig6} shows the Coulomb energy as a function of the total 
angular momentum $L$ for the system of three electrons each with $l=7$,
i.e. at the filling factor $\nu={1\over7}$.
Three frames correspond to the lowest LL (a) and two excited LL's (b,c).
Insets show the spectra for $l=3$ (filling factor $\nu={1\over3}$).
Since individual electron angular momentum $l=S+n$ is the same in all main
frames, the three electron Hilbert spaces contain the same $L$ multiplets.
The difference between spectra (a), (b), and (c) comes from different 
Coulomb matrix elements, i.e.\ different pseudopotentials $V({\cal R})$, 
in different LL's.

For the lowest LL, the Coulomb interaction plotted in 
figure~\ref{fig5}(a) behaves like the hard core repulsion $V_{\rm HC}$ 
defined in equation~\ref{eqvhc}.
The energy spectrum in figure~\ref{fig6}(a) splits into bands of states 
with no parentage from pair states with ${\cal R}<7$ (diamonds), ${\cal 
R}<5$ (open circles), ${\cal R}<3$ (full circles), and remaining states 
with parentage from all pair states including ${\cal R}=1$ (squares).
The CFP's which are expected to vanish identically for any pseudopotential 
(see the last column in table~\ref{tab4}) or which would vanish for the 
eigenstates of the interaction $V_{\rm HC}$ defined in equation~\ref{eqvhc}, 
indeed vanish or are very small (${\cal F}<0.01$) for the eigenstates of 
the Coulomb interaction.
This means that the Coulomb interaction within the lowest LL acts like
$V_{\rm HC}$ and the two interactions have essentially identical 
eigenstates.

Since $V(1)-V(3)>V(3)-V(5)>\dots$ in figure~\ref{fig5}(a), the gap 
between the highest energy band (${\cal R}\ge1$) and the lower ones is 
the largest, the gap below the ${\cal R}\ge3$ band is the next largest,
etc.
The lowest band (${\cal R}\ge7$) consists of only one state at $L=0$.
This is the Laughlin $\nu={1\over7}$ ground state.
The excitation gap above the $\nu={1\over7}$ state is governed by 
$V(5)-V(7)$ and, as might be expected, it is almost unobservable. 
Note also that the first excited band in figure~\ref{fig6}(a) containing
states with ${\cal R}\ge5$ consists of multiplets at $L=2$, 3, 4, and 6, 
in contrast with the mean field CF prediction ($L=1$, 2, and 3).

The inset in figure~\ref{fig6}(a) shows the spectrum for $l=3$.
The $L=0$ ground state has ${\cal F}(1)=0$ (see the first column in 
table~\ref{tab4}); this is the Laughlin $\nu={1\over3}$ state.
The structure of energy spectrum for $l=3$ is very similar to that within 
the two lowest bands for $l=7$.
This is because the Coulomb interaction for $n=0$ acts like hard core 
repulsion and decreasing angular momentum by $p(N-1)$ is equivalent to 
introduction of a hard core which forbids pair states with ${\cal R}<2p+1$
(see figure~\ref{fig7} and the discussion in the following section).

The Coulomb pseudopotentials for $n=0$ in figure~\ref{fig5}(a) and 
for $n=1$ in figure~\ref{fig5}(b) behave similarly for ${\cal R}\ge3$.
In consequence, the two lowest bands of states in figure~\ref{fig6}(a)
and (b) look similar.
The CFP's which are expected to be small, are found to be smaller than
0.01 for both $n=0$ and 1.
However, for the smallest ${\cal R}$, the condition $V(1)-V(3)\gg 
V(3)-V(5)$ is no longer satisfied for $n=1$.
Close to ${\cal R}=1$, the Coulomb pseudopotential for $n=1$ decreases 
too slowly with increasing ${\cal R}$, and its eigenstates, having some
parentage from the ${\cal R}=1$ pair state, are significantly different
from those of the hard core repulsion.
For example, the states at $L=10$ and 12 marked with full dots in 
figure~\ref{fig6}(b) both have significant parentage from ${\cal R}=1$,
${\cal F}(1)\approx0.11$, while the two other states with $L=10$ and 
12, marked with squares, both have ${\cal F}(1)\approx0.23$, only 
twice as large.
For the same reason, there is almost no gap above the ${\cal R}\ge3$ 
band for $n=1$, in contrast to the $n=0$ spectrum.

Different behavior of $V({\cal R})$ for $n=1$ at small values of 
${\cal R}$ has much more pronounced effect on the $l=3$ spectrum shown 
in the inset.
The $L=0$ state must have ${\cal F}(1)=0$ because of the angular 
momentum addition argument (see table~\ref{tab4}), but it is no longer 
the ground state.
Let us stress this result: for three electrons, the Laughlin like 
$\nu={1\over3}$ state is not the ground state in the first excited LL.
Hence, the Laughlin like $\nu=2+{1\over3}$ state is not the ground 
state of the 13 electron system at $2S=4$.
However, the Laughlin like $\nu=2+{1\over7}$ state remains the ground 
state of 29 electrons at $2S=12$.

For $n=2$, the Coulomb pseudopotential in figure~\ref{fig5}(c) 
deviates from that for $n=0$ at all ${\cal R}<5$, and the only gap 
which persists in the spectrum in figure~\ref{fig6} is that above 
the ${\cal R}\ge7$ ground state.
Higher bands, containing states with the smallest appropriate CFP 
(which would be zero for the hard core repulsion) are not even ordered 
as those for $n=0$ or 1.
In the inset, the Laughlin $\nu={1\over3}$ state with ${\cal R}\ge3$ 
is the highest energy state for $n=2$.

\section{Many electron systems}
\label{secVI}

\subsection{Coefficients of fractional grandparentage}
\label{secVIa}

Equations~\ref{eq1} and \ref{eq2} can be generalized to the case of
an arbitrary number of electrons.
An antisymmetric wavefunction $\left|l^N,L\alpha\right>$ of $N$ electrons 
each with angular momentum $l$ that are combined to give a total angular 
momentum $L$ can be written as (Shalit and Talmi 1963, Cowan 1981)
\begin{equation}
   \left|l^N,L\alpha\right>=
   \sum_{L_{12}}\sum_{L'\alpha'} 
   G_{L\alpha,L'\alpha'}(L_{12})
   \left|l^2,L_{12};l^{N-2},L'\alpha';L\right>.
\label{eq3}
\end{equation}
Here, $\left|l^2,L_{12};l^{N-2},L'\alpha';L\right>$ denote product 
states in which angular momenta $l_1=l_2=l$ of two electrons are added 
to obtain pair angular momentum $L_{12}$, then angular momenta $l_3=l_4
=\dots=l_N=l$ of remaining $N-2$ electrons are added to obtain angular 
momentum $L'$ (different states with this angular momentum are labeled 
with different $\alpha'$), and finally $L_{12}$ is added to $L'$ to 
obtain total angular momentum $L$.
The state $\left|l^N,L\alpha\right>$ is totally antisymmetric, while 
states $\left|l^2,L_{12};l^{N-2},L'\alpha';L\right>$ are antisymmetric 
under interchange of particles 1 and 2, and under interchange of any 
pair of particles 3, 4, \dots, $N$.
The factor $G_{L\alpha,L'\alpha'}(L_{12})$, or $G_{L\alpha,L'\alpha'}
({\cal R})$ where ${\cal R}=2l-L_{12}$, is called the coefficient 
of fractional grandparentage (CFGP).
For $N=3$, it is equivalent to the CFP, $G_{L\alpha,l}({\cal R})\equiv 
F_{L\alpha}({\cal R})$.

The two particle interaction matrix element expressed through the 
CFGP's is
\begin{eqnarray}
   \left<l^N,L\alpha\right|&V&\left|l^N,L\beta\right>
   ={N(N-1)\over2}
\nonumber\\
   &\times&
   \sum_{\cal R} \sum_{L'\alpha'} 
   G_{L\alpha,L'\alpha'}({\cal R})\,G_{L\beta,L'\alpha'}({\cal R})\, 
   V({\cal R}).
\end{eqnarray}
For an interaction eigenstate, its energy is
\begin{equation}
   E_{L\alpha}={N(N-1)\over2}
   \sum_{\cal R} {\cal G}_{L\alpha}({\cal R}) \, V({\cal R}),
\label{eq7}
\end{equation}
where the coefficient
\begin{equation}
   {\cal G}_{L\alpha}({\cal R})=
   \sum_{L'\alpha'}\left|G_{L\alpha,L'\alpha'}({\cal R})\right|^2
\label{eq52}
\end{equation}
gives the probability that a pair of electrons $ij$ are in the pair 
state of a given ${\cal R}$.
The derivation of the CFGP's for arbitrary $N$ and $l$ is rather tedious.
Note however that the coefficients ${\cal G}({\cal R})$ can be expressed 
as (compare equation~\ref{eqcalFa})
\begin{equation}
   |{\cal G}_{L\alpha}({\cal R})|^2=
   \left<L\alpha\right|\hat{\cal P}_{ij}({\cal R})\left|L\alpha\right>
\end{equation}
and calculated as the expectation value of the `selective interaction' 
Hamiltonian $\hat{H}_{\cal R}$, whose only nonvanishing pseudopotential 
parameter is $V({\cal R})=1$ (compare equation~\ref{eqcalFb}), 
\begin{equation}
   {\cal G}_{L\alpha}({\cal R})=
   {2\over N(N-1)} 
   \left<L\alpha\right|\hat{H}_{\cal R}\left|L\alpha\right>.
\end{equation}
From the orthonormality of functions $\left|l^N,L\alpha\right>$ it 
is also apparent that 
\begin{equation}
   \sum_{\cal R} \sum_{L'\alpha'}
   G_{L\alpha,L'\alpha'}({\cal R})\, G_{L\beta,L'\alpha'}({\cal R})
   =\delta_{\alpha\beta}.
\label{eq6}
\end{equation}

\subsection{Dynamical symmetry of hard core repulsion}
\label{secVIb}

The angular momentum addition argument fails for more than three 
electrons, and there are no $L$ multiplets for $N>3$ whose CFGP for 
${\cal R}=1$, 3, \dots\ would vanish regardless of the form of 
interaction pseudopotential.
However, the many electron Hilbert space ${\cal H}$ still contains 
subspaces ${\cal H}_p$ holding many body states with grandparentage 
only from pair states with ${\cal R}\ge2p+1$, for which ${\cal G}(1)=
{\cal G}(3)=\dots={\cal G}(2p-1)=0$,
\begin{equation}
   {\cal H}\equiv{\cal H}_0\supset{\cal H}_1
                           \supset{\cal H}_2\supset\dots
\end{equation}
The total Hilbert space splits thus into subspaces $\tilde{\cal H}_p=
{\cal H}_p\setminus{\cal H}_{p+1}$, containing many body states that do 
not have grandparentage from pair states with ${\cal R}<2p+1$, but have 
some grandparentage from ${\cal R}=2p+1$,
\begin{equation}
   {\cal H}=\tilde{\cal H}_0\oplus\tilde{\cal H}_1\oplus
            \tilde{\cal H}_2\oplus\dots
\end{equation}
For $N$ electrons on a Haldane sphere each with angular momentum $l$,
there is more than one subspace (subspace $\tilde{\cal H}_1$ is not 
empty) for $2l\ge3(N-1)$, i.e.\ for filling factors $\nu\le{1\over3}$.
In general, $\tilde{\cal H}_p$ is not empty (some states with 
${\cal R}\ge2p+1$ can be constructed) for $\nu\le(2p+1)^{-1}$.

The subspaces $\tilde{\cal H}_p$ are the eigensubspaces of the hard 
core repulsive potential $V_{\rm HC}$ defined in equation~\ref{eqvhc}, 
whose low energy states have to avoid grandparentage from pair states 
with large repulsion (small ${\cal R}$).
Consequently, as for three electrons, the energy levels in the many 
electron spectrum with hard core interaction form bands corresponding 
to subspaces $\tilde{\cal H}_p$.
For given $N$ and $l$, i.e.\ for a given filling factor $\nu$ such that 
$(2p+3)^{-1}<\nu\le(2p+1)^{-1}$, there are $(p+1)$ bands, and the $q$th 
band ($q=0$, 1, \dots, $p$) corresponds to $\tilde{\cal H}_q$.
The $p$th band is the lowest energy band with the maximum number of 
CFGP's vanishing, and the 0th band is the highest energy band containing 
states with some grandparentage from the ${\cal R}=1$ pair state.
The energy gap between the $q$th band and the $(q+1)$-st band is of the 
order of $V(2q+1)-V(2q+3)$.
Hence, the largest gap is that between the 0th band and the 1st band, 
the next largest is that between the 1st band and 2nd band, etc.

\begin{figure}[t]
\epsfxsize=3.4in
\epsffile{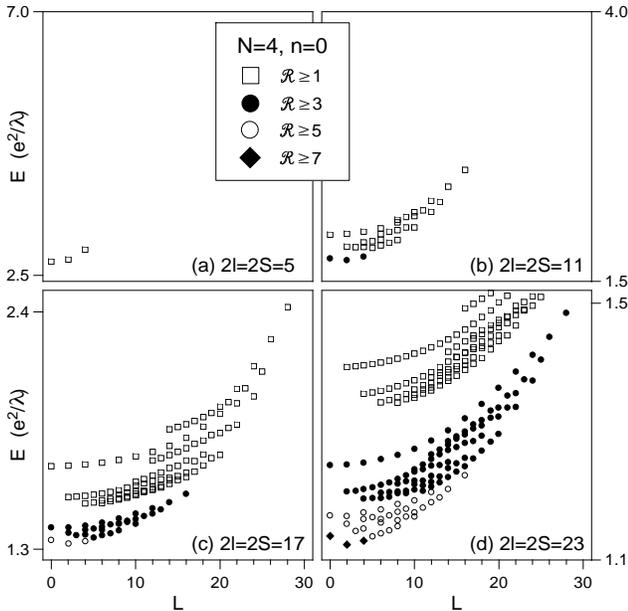}
\caption{
   The energy spectra of four electrons in the lowest Landau level at 
   different monopole strength: $2S=5$ (a), 11 (b), 17 (c), and 23 (d).
   All those $2S$ values are equivalent in the mean field composite 
   Fermion picture (Chern--Simons transformation with $p=0$, 1, 2, 
   and 3, respectively).
   Diamonds: states with ${\cal R}\ge7$, 
   i.e.\ ${\cal G}(1)\approx{\cal G}(3)\approx{\cal G}(5)\approx0$ 
   and ${\cal G}(7)>0$; 
   open circles: ${\cal R}\ge5$, 
   i.e.\ ${\cal G}(1)\approx{\cal G}(3)\approx0$ and ${\cal G}(5)>0$; 
   full circles: ${\cal R}\ge3$, 
   i.e.\ ${\cal G}(1)\approx0$ and ${\cal G}(3)>0$; and
   open squares: ${\cal R}\ge1$, 
   i.e.\ ${\cal G}(1)>0$.}
\label{fig7}
\end{figure}

Importantly, the set of angular momentum multiplets which make the $q$th 
band ($\tilde{\cal H}_q$ subspace) of the spectrum of $N$ electrons 
each with angular momentum $l$ is always the same as the set of multiplets 
in the $(q+1)$-st band ($\tilde{\cal H}_{q+1}$ subspace) of $N$ electrons 
each with angular momentum $l+(N-1)$.
This is demonstrated in figure~\ref{fig7} for four electrons in the
lowest LL interacting through the (hard core like) Coulomb pseudopotential.
When $l=S$ is increased by $N-1$, the only significant difference in the 
spectrum is the appearance of an additional band at high energy.
The structure of the low energy part of the spectrum is completely 
unchanged.
All bands and multiplets in the spectrum for $2S$ correspond directly 
to appropriate bands and multiplets in the spectrum for the monopole 
strength $2S+2(N-1)$.
For example, all three allowed multiplets at $2S=5$ ($L=0$, 2, and 4)
form the low energy band at $2S=11$, 17, and 23, where they span the
$\tilde{\cal H}_1$, $\tilde{\cal H}_2$ and $\tilde{\cal H}_3$ subspaces,
respectively.
Similarly, the first excited band at $2S=11$ (open squares in frame b) 
is repeated in the spectra for $2S=17$ and 23, where it corresponds to 
$\tilde{\cal H}_1$ and $\tilde{\cal H}_2$ subspace, respectively.

Let us repeat that the fact that identical multiplets occur in subspace 
$\tilde{\cal H}_q$ for single electron angular momentum $l$, and in 
subspace $\tilde{\cal H}_{q+1}$ for $l$ replaced by $l+(N-1)$, does not 
depend on the form of interaction, and follows solely from the rules of 
addition of angular momenta of identical Fermions.
However, if the interaction pseudopotential has the hard core properties 
as in equation~\ref{eqvhc}, then the many body interaction Hamiltonian 
has a new, dynamical symmetry, as a result of which:
(i) the subspaces $\tilde{\cal H}_q$ are the eigensubspaces and the 
subspace (band) index $q$ is a good quantum number;
(ii) the energy bands corresponding to $\tilde{\cal H}_q$ with higher 
$q$ lie below those of lower $q$; 
(iii) the spacing between neighboring bands is governed by a difference 
between appropriate pseudopotential coefficients; and
(iv) the wavefunctions and the relative position of energy levels 
within each ($q$th) band do not depend on the details of interaction 
(it will be shown later that they repeat the spectrum of 
${\cal G}(2q+1)$; see figure~\ref{fig15}).
Replacing the model hard core pseudopotential by a `softer' one (the 
measure of the 'hard core' character $\beta$ will be specified in 
section~\ref{secVIIe}) leads to:
(i) coupling between subspaces $\tilde{\cal H}_q$;
(ii) mixing, overlap, or even order reversal of bands;
(iii) deviation of wavefunctions and the structure of energy levels 
within bands from those of the hard core repulsion (and thus their
dependence on details of the interaction pseudopotential).

The reoccurrence of $L$ multiplets forming the low energy band when 
$l$ is replaced by $l\pm(N-1)$ has the following crucial implication.
The lowest energy, $p$th band contains $L$ multiplets which are all 
the allowed multiplets of $N$ electrons each with angular momentum 
$l-p(N-1)$.
This is because if $(2p+3)^{-1}<\nu_{N,l}\le(2p+1)^{-1}$, then 
${1\over3}<\nu_{N,l-p(N-1)}\le1$ and there is only one, 0th band in 
the spectrum.
As for three electrons, for the lowest LL with $l=S$ this means that 
the lowest energy band at the monopole strength $2S$ contains a subset 
of low energy multiplets which are all the allowed multiplets at 
a smaller monopole strength $2S-2p(N-1)$.
But $2S-2p(N-1)$ is just $2S^*$, the effective monopole strength of CF's!
The mean field CS transformation which attaches $2p$ fluxes (vortices) to 
each electron selects the same $L$ multiplets from the entire spectrum as 
does the introduction of a hard core, which forbids a pair of electrons 
to be in a state with ${\cal R}<2p+1$.

The success of the mean field CF picture in prediction of the low lying 
band of states in the many electron spectrum relies on the fact that the
Coulomb interaction within the lowest LL acts like the hard core repulsion.
For filling factors $\nu$ such that $(2p+3)^{-1}<\nu\le(2p+1)^{-1}$, 
the states predicted by the mean field CF picture as the states of an 
appropriate number of QH's in the Laughlin $\nu=(2p+1)^{-1}$ ground 
state are the states which for the hard core interaction have the 
maximum number ($p$) of vanishing CFGP's associated with the highest 
pseudopotential parameters.
These are the states with ${\cal R}\ge2p+1$ spanning subspace 
$\tilde{\cal H}_p$.
In particular, there is always only one state with ${\cal R}\ge2p+1$ 
($\tilde{\cal H}_p$ is one dimensional) at the filling factor 
$\nu=(2p+1)^{-1}$.
This state has $L=0$ and it is the Laughlin incompressible ground state, 
separated from other states by the gap $\Delta$ which is of the order 
of $\Delta=V(2p-1)-V(2p+1)$.

As long as the eigenstates of the Coulomb interaction are approximately
those of the hard core repulsive interaction, the incompressible ground
states are associated with appearance of states with significantly lower 
CFGP's than all other states in the spectrum.
The Laughlin $\nu=(2p+1)^{-1}$ ground states are the only states with 
${\cal G}(1)\approx{\cal G}(3)\approx\dots\approx{\cal G}(2p-1)\approx0$
in their Hilbert spaces (the CFGP's do not vanish identically due to the 
weak mixing between $\tilde{\cal H}_q$ subspaces).
All other states have some (significant) grandparentage from pair states 
with ${\cal R}<2p+1$.
The Jain states at filling factors $\nu$ in the range $(2p+3)^{-1}<\nu<
(2p+1)^{-1}$ are those of all states with ${\cal G}(1)\approx{\cal G}(3)
\approx\dots\approx{\cal G}(2p-1)\approx0$, for which ${\cal G}(2p+1)$, 
the first nonvanishing CFGP, is significantly smaller than for other 
states (W\'ojs and Quinn 1999a).

What is the condition for the interaction pseudopotential to behave
like the hard core repulsion and have the energy spectrum characteristic 
of the FQH effect?
In the following sections we answer this question and explain why the hard 
core type (FQH) ground states occur for the Coulomb interaction within the 
lowest LL.
We also show that due to a different form of the Coulomb pseudopotential 
in higher (spin polarized) LL's, the FQH ground states for $n>0$ occur 
only at lower densities, when, at low energy, only the hard core like 
part of the pseudopotential (at high ${\cal R}$) contributes to the 
Hamiltonian given by equation~\ref{hproj}.

\subsection{Coulomb interaction in lowest and excited Landau levels}
\label{secVIc}

\begin{figure}[t]
\epsfxsize=3.4in
\epsffile{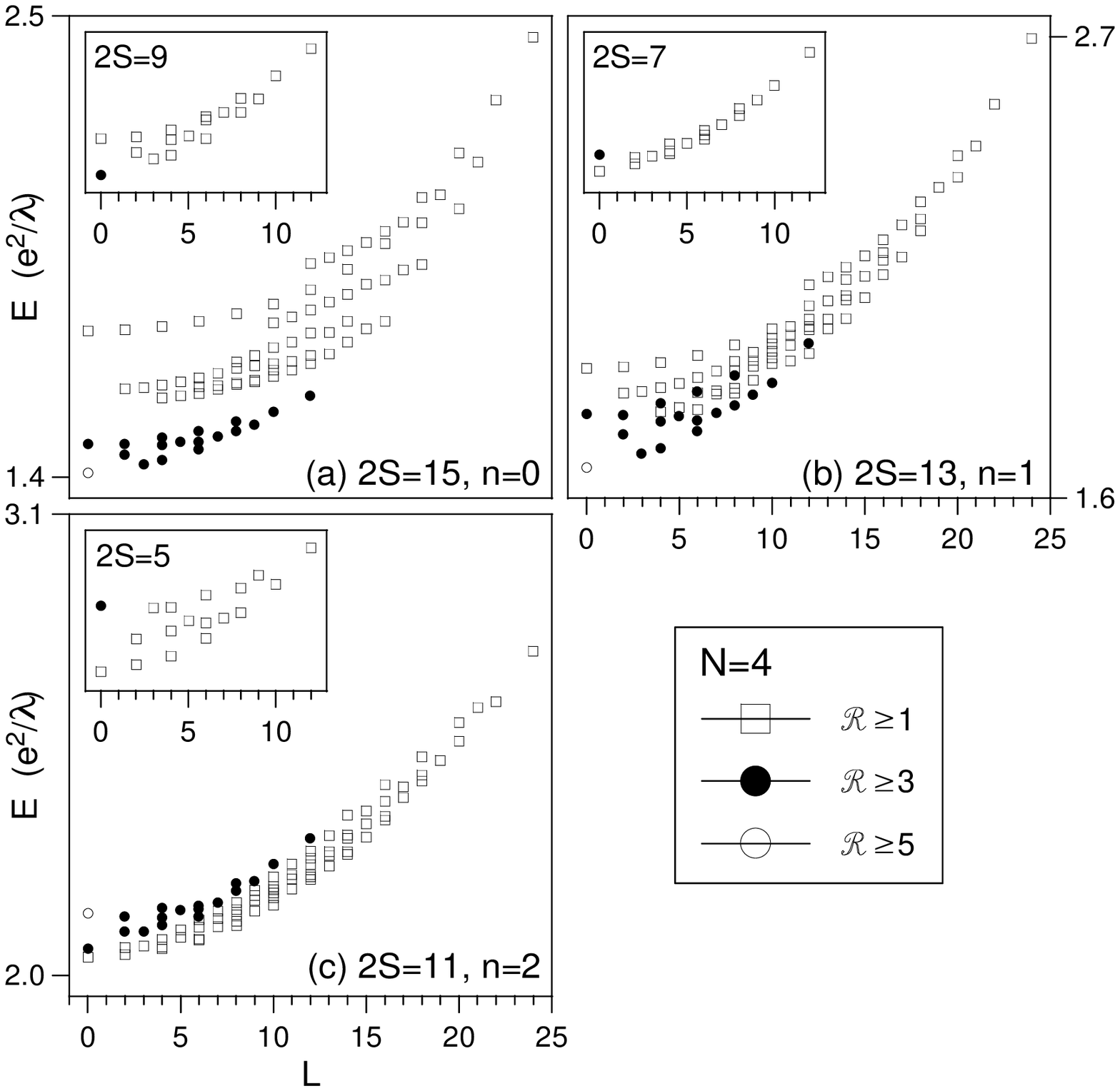}
\caption{
   The Coulomb energy of four electrons each with $l={15\over2}$ in the 
   lowest (a), first excited (b), and second excited (c) Landau level.
   Open circles: states with ${\cal R}\ge5$, 
   i.e.\ ${\cal G}(1)\approx{\cal G}(3)\approx0$ and ${\cal G}(5)>0$; 
   full circles: ${\cal R}\ge3$, 
   i.e.\ ${\cal G}(1)\approx0$ and ${\cal G}(3)>0$; and
   squares: ${\cal R}\ge1$, 
   i.e.\ ${\cal G}(1)>0$.
   The ground states for $n=0$ and 1 are the Laughlin $\nu={1\over5}$ 
   states within these LL's.
   Insets: spectra for $l={9\over2}$; 
   the ground state for $n=0$ is the Laughlin $\nu={1\over3}$ state.}
\label{fig8}
\end{figure}

Figure~\ref{fig8} shows the Coulomb energy as a function of $L$ for the 
system of four electrons each with $l={15\over2}$.
Three frames correspond to the lowest LL ($n=0$) and two excited LL's 
($n=1$ and 2), and the insets show the spectra for $l={9\over2}$.
Figure~\ref{fig8} is very similar to figure~\ref{fig6}, and demonstrates 
that the conclusions drawn for the simple three electron system remain 
valid for an arbitrary $N$.

As for three electrons, the Coulomb interaction within the lowest LL 
($n=0$) behaves like the hard core interaction and the energy spectrum 
splits into bands of states with ${\cal R}\ge5$ (open circles), ${\cal 
R}\ge3$ (full circles), and ${\cal R}\ge1$ (squares).
The ${\cal R}\ge5$ band contains only the Laughlin $\nu={1\over5}$ 
ground state.
For $N>3$, none of CFP's vanish identically for an arbitrary interaction,
but the CFP's which would vanish for the eigenstates of the hard core 
interaction defined in equation~\ref{eqvhc}, indeed vanish or are very 
small (${\cal G}<0.01$) for the eigenstates of the Coulomb interaction.
The inset in figure~\ref{fig5}(a) shows the spectrum for $l={9\over2}$, 
with the Laughlin $\nu={1\over3}$ ground state.
The energy spectrum for $l={9\over2}$ repeats main features of the two 
lowest energy bands for $l={15\over2}$.

Within the first excited LL ($n=1$), only the lowest band with ${\cal 
R}\ge5$ can be distinguished.
The two higher bands (${\cal R}\ge3$ and ${\cal R}\ge1$) overlap.
Also, some of the coefficients ${\cal G}(1)$ which would be zero for the 
hard core repulsion, are quite large ($>0.1$) for $n=1$.
In the inset, the two $L=0$ states have ${\cal G}(1)=0.08$ and 0.26, 
and the Laughlin like $\nu={1\over3}$ state with a smaller ${\cal G}(1)$ 
(full circle) is the one with higher energy.
Even though the ground state has $L=0$, it is not the state with the 
Laughlin like correlations, with electrons avoiding pair states with
the largest repulsion (i.e.\ smallest average separation, see 
equation~\ref{eqharm}).
The gap above this ground state is not associated with the energy 
$V(1)-V(3)$, and hence the $\nu=2+{1\over3}$ state is unlikely to be 
an incompressible ground state in the thermodynamic limit.

For $n=2$, neither the Laughlin like $\nu={1\over5}$ state in the 
main frame (${\cal R}\ge5$, open circle), nor the Laughlin like 
$\nu={1\over3}$ state in the inset (${\cal R}\ge3$, full circle) is 
the ground state.
This suggests that neither the $\nu=4+{1\over3}$ state nor the 
$\nu=4+{1\over5}$ state is an incompressible ground state in the 
thermodynamic limit.

We have calculated the energy spectra analogous to those in 
figure~\ref{fig8} for different numbers of electrons and conclude that 
the Laughlin like $L=0$ state with $\nu={1\over3}$, which is the only 
state with ${\cal R}\ge3$ in its spectrum, is the ground state only 
within the lowest LL ($n=0$).
Similarly, the Laughlin like $\nu={1\over5}$ state with ${\cal R}\ge5$ 
is the ground state only for $n\le1$.

\begin{figure}[t]
\epsfxsize=3.4in
\epsffile{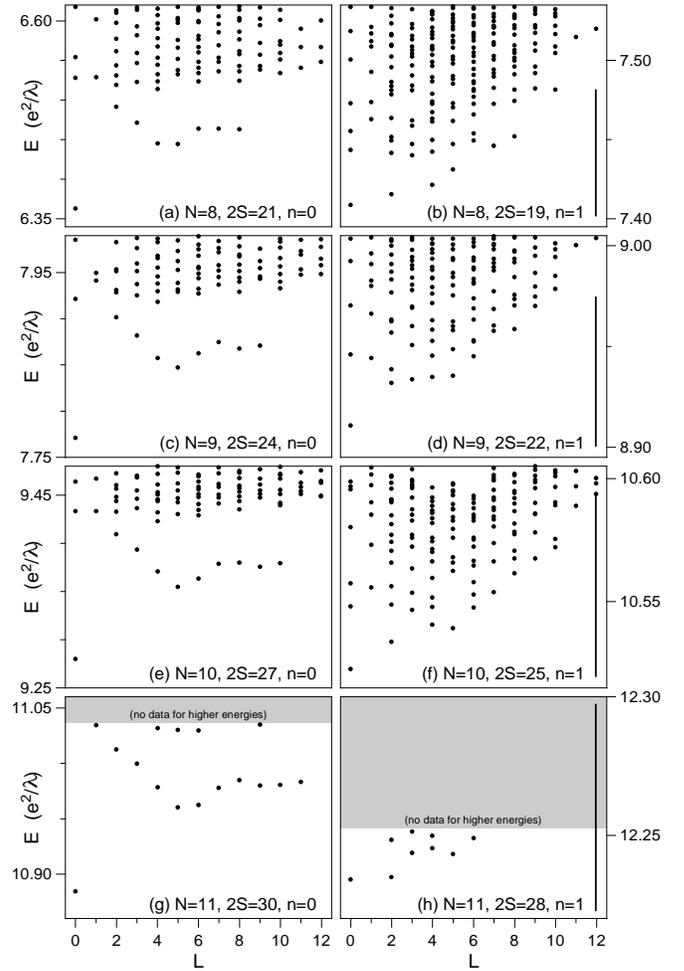}
\caption{
   The energy spectra of eight (top) to eleven (bottom) electrons 
   in the lowest (left) and first excited (right) Landau level at 
   the filling factor $\nu={1\over3}$.}
\label{fig9}
\end{figure}

\begin{figure}[t]
\epsfxsize=3.4in
\epsffile{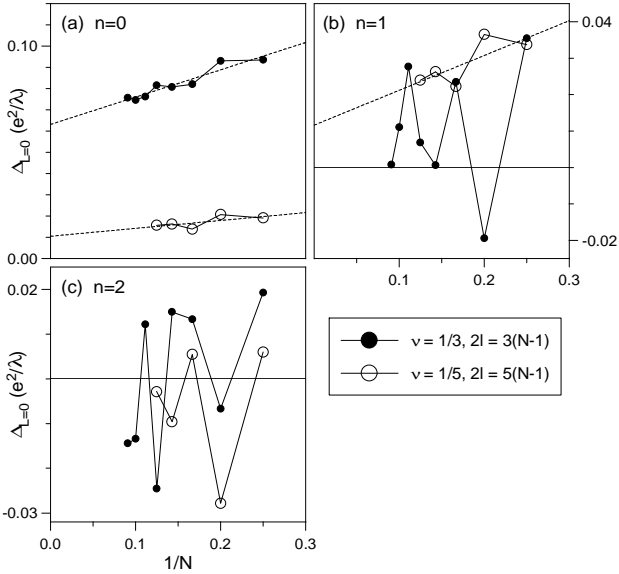}
\caption{
   The energy gap $\Delta_{L=0}$ from the lowest $L=0$ state to the 
   lowest state at $L>0$ as a function of the inverse electron number
   $N^{-1}$, for the lowest (a), first excited (b),
   and second excited (c) Landau level.
   Full circles: $\nu=2n+{1\over3}$;
   open circles: $\nu=2n+{1\over5}$.
   The dashed lines give linear fits for the Laughlin like 
   incompressible ground states at $\nu={1\over3}$, $\nu={1\over5}$, 
   and $\nu=2+{1\over5}$.
   The ground states at $\nu=2+{1\over3}$, $4+{1\over3}$, and 
   $4+{1\over5}$ are unlikely to be incompressible for 
   $N\rightarrow\infty$.}
\label{fig10}
\end{figure}

The angular momentum $L$ of the ground state of $N$ electrons at the
monopole strength $2S$ corresponding to the $\nu={1\over3}$ filling 
within the LL of $n>0$ or to the $\nu={1\over5}$ filling within the LL 
of $n>1$ depends on $N$.
Even though $L=0$ (ground state is nondegenerate) for some values 
of $N$, the low lying spectra do not resemble those in the lowest LL, 
and the excitation is not associated with energy $V(1)-V(3)$.
In order to verify if the $L=0$ ground states with $\nu=2+{1\over3}$,
$2+{1\over5}$, $4+{1\over3}$, and $4+{1\over5}$ are incompressible 
ground states in the thermodynamic limit, we have calculated the energy 
gaps above these states for different values of $N$.
The energy spectra of up to eleven electrons at the filling factor 
$\nu={1\over3}$ in the lowest and first excited LL's are presented in 
figure~\ref{fig9}.
The energy scales for $n=0$ and 1 are different, and the bar in the bottom 
right corner of each $n=1$ graph on the right shows the energy gap of 
the corresponding system in the lowest LL on the left.
Figure~\ref{fig10} shows the dependence of the gap $\Delta_{L=0}$ from 
the lowest $L=0$ state to the lowest state of $L>0$, as a function of 
$N^{-1}$.
For filling factors $\nu=2n+{1\over3}$ (full circles), $N$ varies 
between four and eleven, and for $\nu=2n+{1\over5}$ (open circles) 
$N$ goes up to eight.
Negative $\Delta_{L=0}$ means that the ground state is degenerate 
(has $L>0$).
In such case, $|\Delta_{L=0}|$ gives the excitation energy from
this degenerate ground state to the lowest state at $L=0$.

For $n=0$, the ground states at both $\nu={1\over3}$ and ${1\over5}$ 
are Laughlin incompressible states.
The gap $\Delta$ persists for $N\rightarrow\infty$, and the estimates 
obtained from the best linear fits (dashed lines) are $\Delta_{\nu=1/3}
=0.0632\,e^2/\lambda$ and $\Delta_{\nu=1/5}=0.0105\,e^2/\lambda$.
For $n=1$, the $L=0$ state at $\nu=2+{1\over5}$ is the Laughlin like 
state and the gap above it seems to converge to a finite value; the 
linear fit gives $\Delta_{\nu=2+1/5}=0.0116\,e^2/\lambda$, very close 
to $\Delta_{\nu=1/5}$.
On the other hand, the dependence of the gap $\Delta$ above the (non 
Laughlin like) states at $\nu=2+{1\over3}$, $4+{1\over3}$, and 
$4+{1\over5}$ on the electron number $N$ is quite different than 
those for Laughlin states.
No conclusive statement about the incompressibility (or even the sign 
of $\Delta$, i.e.\ the nondegeneracy) of these states in the thermodynamic 
limit can be made based our finite-size calculations for up to eleven
electrons.
Since at least at $\nu=2+{1\over3}$ the FQH plateau has been observed 
experimentally (Willet {\sl et al.}\ 1987), we have to restrict 
ourselves to repeating the statement (MacDonald and Girvin 1986) 
that the nature of the low lying states at $\nu=2+{1\over3}$, 
$4+{1\over3}$, and $4+{1\over5}$ is different than of the Laughlin 
$\nu={1\over3}$ and ${1\over5}$ states.
In general, low lying states in the lowest and $n$th LL's have Lauglin 
like correlations only below the filling factor $\nu=(2n+1)^{-1}$.
At fillings $\nu\ge(2n+1)^{-1}$ in the $n$th LL, the correlations are
different, possible incompressibility has a different origin, the 
excitation gap is not simply related to the difference between 
appropriate pseudopotential parameters, and the excitations do not
contain Laughlin QP's.

\begin{figure}[t]
\epsfxsize=3.4in
\epsffile{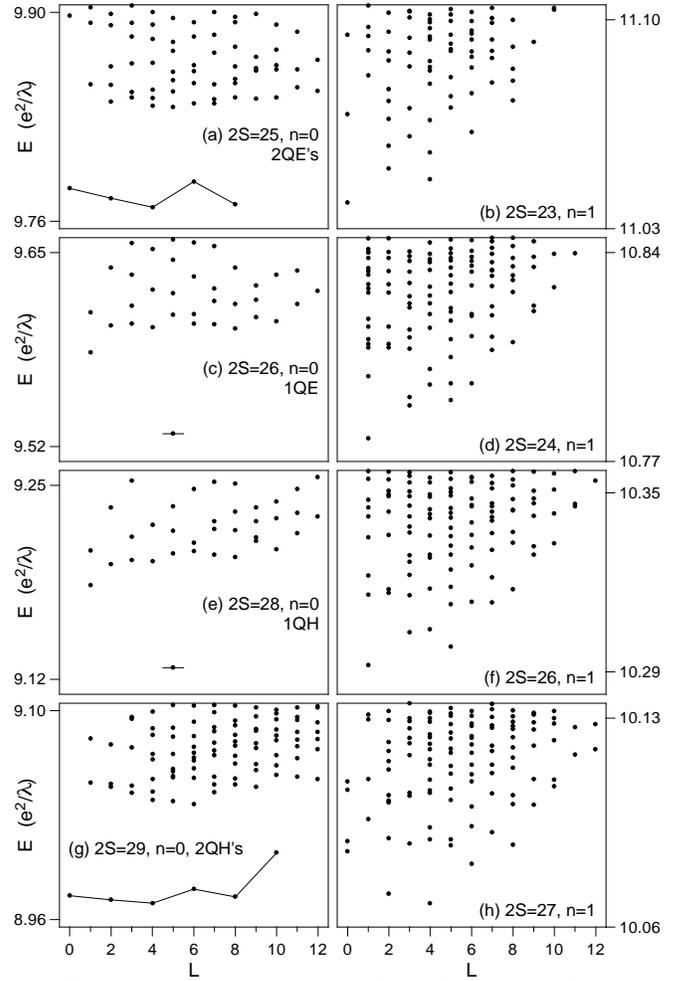}
\caption{
   The energy spectra of ten electrons in the lowest (left) and first 
   excited (right) Landau level at filling factors near $\nu={1\over3}$.}
\label{fig11}
\end{figure}

A clear signature of the non Laughlin like character of the $n={1\over3}$ 
state in excited LL's is the lack of QP type excitations at neighboring 
filling factors.
In figure~\ref{fig11} we compare the energy spectra of ten electrons at 
equal fillings (near $\nu={1\over3}$) of the lowest and first excited LL's.
In the lowest LL, lowest energy states (marked with lines) contain 
two QE's (a), one QE (c), one QH (e) and two QH's (g) in the Laughlin 
$\nu={1\over3}$ state, while in the first excited LL no similar low lying 
states occur (note also that the energy axes in $n=1$ frames are 
streched by a factor of two compared to the $n=0$ ones).
Note also that the energies connected with lines in figure~\ref{fig11}(a) 
and (g) define the pseudopotentials of a pair of appropriate interacting 
QP's in the Laughlin $\nu={1\over3}$ state.

For a complete report of our numerical results for the lowest LL, let 
us add a few numbers to the tables published earlier (Fano {\sl et al.}\ 
1986).
In table~\ref{tab5} we list the Laughlin ground state energy per particle 
(calculated including interaction with a charge compensating background, 
$-N^2e^2/2R$), the angular momentum and excitation energy of the 
magnetoroton minimum, and the `proper' QE and QH energies (calculated 
including additional fractional charge $\pm e/m$ in the background; 
Haldane and Rezayi 1985a, Fano {\sl et al.}\ 1986), for $N=10$ and 11 
electrons at filling factor $\nu={1\over3}$ and for $N=7$ and 8 electrons 
at $\nu={1\over5}$. 
\begin{table}
\caption{
   The ground energy per particle $E/N$ of the Laughlin ground state, 
   the angular momentum $L$ and excitation energy $\Delta$ of the 
   magnetoroton minimum, and the proper quasielectron and quasihole 
   energies, $\epsilon_{\rm QE}$ and $\epsilon_{\rm QH}$, for $N$ 
   electrons at a filling factor $\nu$.}
\begin{tabular}{ccccccc}
  $\nu$ & 
  $N$ & 
  $E/N$ & 
  $L$ & 
  $\Delta$ & 
  $\epsilon_{\rm QE}$ &
  $\epsilon_{\rm QH}$ \\ 
\hline
  1/3 &  10    & $-0.432841$ & 5 & 0.074715 & 0.085675 & 0.030501 \\
      &  11    & $-0.430623$ & 5 & 0.075706 & 0.084658 & 0.030092 \\
      &$\infty$& $-0.415948$ &---& 0.063177 & 0.073724 & 0.025813 \\
\hline
  1/5 &   7    & $-0.353494$ & 4 & 0.016245 & 0.020188 & 0.009068 \\
      &   8    & $-0.350066$ & 5 & 0.015572 & 0.019278 & 0.008510 \\
      &$\infty$& $-0.332850$ &---& 0.010516 & 0.014912 & 0.006288
\end{tabular}
\label{tab5}
\end{table}
The limiting values for $N\rightarrow\infty$ have been calculated using 
data for these and smaller values of $N$.
For example, the QE and QH energies agree very well with extrapolation 
of the Monte Carlo results in disk geometry: $\varepsilon_{\rm QE}=0.073$ 
and $\varepsilon_{\rm QH}=0.0268$ (Morf and Halperin 1986).

\begin{figure}[t]
\epsfxsize=3.4in
\epsffile{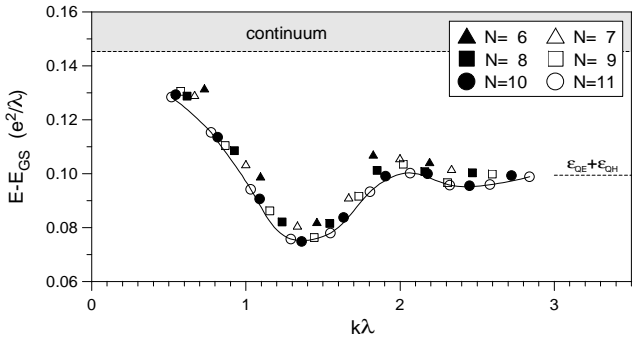}
\caption{
   The excitation energy $E-E_{\rm GS}$ as a function of wavevector $k$
   for the low lying excitations of the Laughlin $\nu={1\over3}$ ground 
   state of six to eleven electrons.}
\label{fig12}
\end{figure}

It is known (Haldane and Rezayi 1985a) that the QE--QH excitonic energy
dispersion (QE--QH pseudopotential) in a Laughlin state, calculated for 
a finite $N$ electron system and plotted as a function of wavevector 
$k=L/R$, quickly converges to the continuous curve of an infinite 2DEG, 
with a magnetoroton minimum at $k$ of the order of the inverse magnetic 
length, $\lambda^{-1}$.
In figure~\ref{fig12} we present the QE--QH dispersion for the 
$\nu={1\over3}$ state, including data for up to eleven electrons.
The continuum marked with a shaded rectangle starts at the lowest 
excitation energy of eleven electrons above the magnetoroton curve.
The $\varepsilon_{\rm QE}+\varepsilon_{\rm QH}=0.099492$ energy 
(our thermodynamic limit estimate) gives the energy of a QE--QP pair
at an infinite distance (infinite $k$).
The smooth solid curve connects data points for $N=11$.

\subsection{Grandparentage coefficients of low lying states}
\label{secVId}

\begin{figure}[t]
\epsfxsize=3.4in
\epsffile{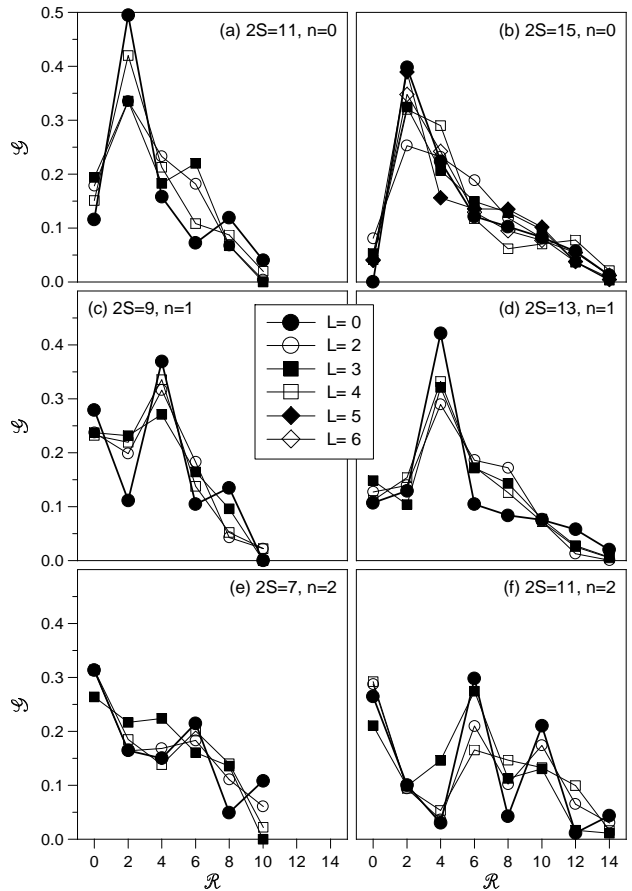}
\caption{
   The grandparentage coefficients ${\cal G}_{L\alpha}({\cal R})$ as 
   a function of relative pair angular momentum ${\cal R}$ for the 
   lowest energy multiplets of six electrons each with $l={11\over2}$ 
   (left) and $l={15\over2}$ (right), calculated for the lowest (top),
   first excited (middle), and second excited (bottom) Landau level.}
\label{fig13}
\end{figure}

Typical dependences of ${\cal G}_{L\alpha}$ on ${\cal R}$ for low lying 
states are plotted in figure~\ref{fig13} for a six electron system at 
$l={11\over2}$ ($\nu={2\over5}$) and $l={15\over2}$ ($\nu={1\over3}$), 
in the lowest and first two excited LL's.
In each frame, thicker line and larger symbols (dots) mark data 
corresponding to the state at $L=0$.
The CFGP profile ${\cal G}({\cal R})$ can be regarded as a pair 
correlation function, except that the probability ${\cal G}$ is given 
as a function of a pair quantum number ${\cal R}$ rather than of 
a pair distance.

In figure~\ref{fig13}(a), the $L=0$ state is the Jain $\nu={2\over5}$ 
ground state and the states with $L=2$, 3, and 4 contain a single 
QE--QH pair. 
Similarly, in figure~\ref{fig13}(b), the $L=0$ state is the Laughlin 
$\nu={1\over3}$ ground state and the states of a single QE--QH pair 
have $L=2$, 3, 4, 5, and 6.
Typically for the low energy states in the lowest LL (or for any 
other short range interaction pseudopotential) at $\nu\ge{1\over3}$, 
${\cal G}(1)$ is small, ${\cal G}(3)$ is large, and for higher ${\cal 
R}$, ${\cal G}$ decreases when ${\cal R}$ increases up to the maximum 
allowed value.
The Jain incompressible ground states always have ${\cal G}(1)$ smaller 
than all other states (by at least 0.035 for $N=6$ and $\nu={2\over5}$).
For Laughlin states, ${\cal G}(1)$ is always negligible (less than 
0.0008 for $N=6$ and $\nu={1\over3}$).
The strong maximum of ${\cal G}({\cal R})$ at ${\cal R}=3$ means that 
a large number of pairs are in the `$\nu={1\over3}$' pair state, on 
a plane given by the Laughlin correlation factor $(z_1-z_2)^3$.

In higher LL's, the ${\cal G}_{L\alpha}({\cal R})$ profiles in 
figure~\ref{fig13}(c--f) differ from those in the lowest LL, but 
they are rather similar for different fillings ($\nu={2\over5}$ and 
${1\over3}$).
Clearly, at any filling or $n$, the low lying states must maximally 
avoid parentage from pair states of highest repulsion.
However, because the pseudopotential $V({\cal R})$ in higher LL's does 
not increase sufficiently quickly with decreasing ${\cal R}$ in its 
entire range, it appears energetically favorable to minimize total 
parentage from a number of pair states with lowest ${\cal R}$, rather 
from a single highest energy state with ${\cal R}=1$.
It appears that requirement of having small total parentage from a 
number of pair states of smallest ${\cal R}$ (smallest separation) 
rather than from a single pair state at ${\cal R}=1$ for a density 
at which only one pair state can be completely avoided implies 
having large parentage from the ${\cal R}=1$ state.
As a result, the maximum of ${\cal G}({\cal R})$ shifts from ${\cal 
R}=3$ (for $n=0$) to ${\cal R}=5$ (for $n=1$) or ${\cal R}=7$ (for 
$n=2$).
Similarly, the minimum at ${\cal R}=1$ (for $n=0$) shifts to ${\cal R}
=3$ (for $n=1$) or ${\cal R}=5$ (for $n=2$).
The occurrence of a large number of pairs in certain pair states
of small ${\cal R}$ (at certain small average distance) and avoiding 
others defines a different type of short range correlation in the 
$\nu={2\over5}$ or $\nu={1\over3}$ states in higher LL's.
The natural interpretation of the maximum at ${\cal R}=1$ for $n>0$ 
instead of the strong minimum as for $n=0$ seems to be the formation 
of electron pairs (Haldane and Rezayi 1988, Moore and Read 1991).
Since the electron--electron interaction is repulsive, the formation
of such pairs is a many body phenomenon and the stability of a pair
requires the presence of a surrounding electron gas at an appropriate
density.
For a given pseudopotential, the pairs could be formed if putting two 
electrons in a pair state with strong repulsion greatly reduces their 
interaction with other electron pairs.
As a result, the gain in total interaction energy in equation~\ref{eq7}
due to reducing the contribution from pair states of intermediate 
${\cal R}$ can exceed the cost due to creating relatively few ($\sim 
N/2$) pairs of the smallest ${\cal R}$.

\begin{figure}[t]
\epsfxsize=3.4in
\epsffile{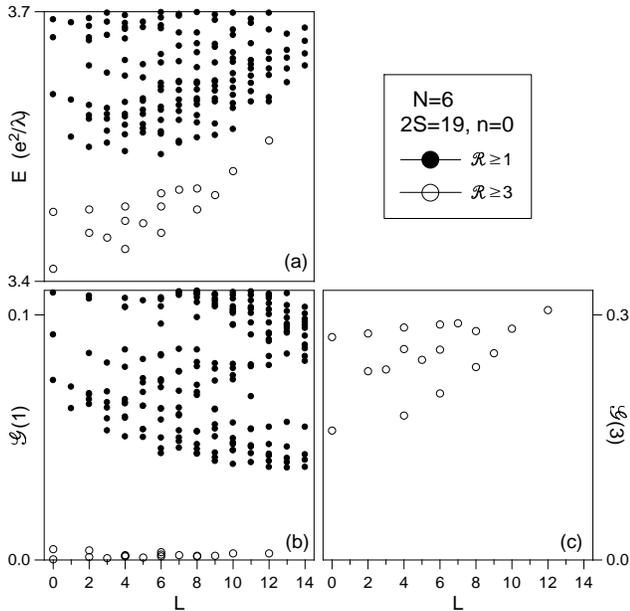}
\caption{
   The energy $E$ (a), and grandparentage coefficients ${\cal G}(1)$ (b), 
   and ${\cal G}(3)$ (c), as a function of angular momentum $L$ for the 
   system of six electrons in the lowest Landau level at $2S=19$.
   Open circles: states with ${\cal R}\ge3$, 
   i.e.\ ${\cal G}(1)\approx0$ and ${\cal G}(3)>0$;
   full circles: states with ${\cal R}\ge1$, 
   i.e.\ ${\cal G}(1)>0$.}
\label{fig14}
\end{figure}

At the values of ${\cal R}$ at which the pseudopotential $V({\cal R})$
decreases very quickly with increasing ${\cal R}$, $V({\cal R})$ is said 
to have short range.
At a given filling factor $\nu$, a number of pair states with largest 
repulsion are avoided completely, and the dominant contribution to 
the energy is the largest term in equation~\ref{eq7}.
This term is the one at the smallest value of ${\cal R}$, for which 
${\cal G}({\cal R})$ does not vanish.
There is a strong correlation between energy and the lowest order 
nonvanishing CFGP, ${\cal G}(2p+1)$ for $(2p+3)^{-1}<\nu\le(2p+1)^{-1}$.
The low energy states always have significantly smaller ${\cal G}(2p+1)$ 
than all other states with ${\cal R}\ge2p+1$.
As an example, in figure~\ref{fig14} we plot energies and coefficients 
${\cal G}(1)$ and ${\cal G}(3)$ for the eigenstates of six electrons 
in the lowest LL at $2S=19$.
The band of multiplets marked with open circles have ${\cal G}(1)<0.0043$, 
and all other states have ${\cal G}(1)>0.037$. 
The energy gap between the two bands in frame (a) is the result of the 
CFGP gap in frame (b).
The states with ${\cal G}(1)\approx0$, are approximate zero energy 
eigenstates of the hard core pseudopotential with $V(1)>0$ and 
all other parameters vanishing.
In the mean field CF picture, these states contain four QH's in the 
Laughlin $\nu={1\over3}$ state, each with angular momentum 
$l_{\rm QH}={9\over2}$.
The angular momentum dependence of energy within this band in frame (a) 
is very similar to that of ${\cal G}(3)$ in frame (c).
In particular, the $L=0$ ground state, which is the $\nu={2\over7}$ 
Jain state in the mean field CF picture, has the lowest ${\cal G}(3)$ of 
all states in this band.

\begin{figure}[t]
\epsfxsize=3.4in
\epsffile{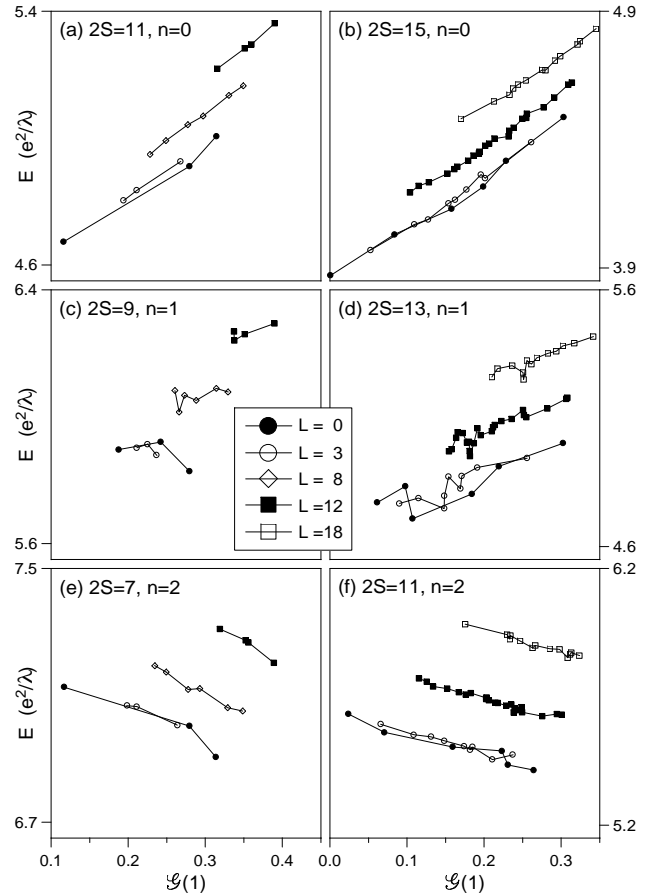}
\caption{
   The Coulomb energy of six electrons each with $l={11\over2}$ (left) 
   and $l={15\over2}$ (right) as a function of the grandparentage 
   coefficient ${\cal G}(1)$, calculated for the lowest (top), first 
   excited (middle), and second excited (bottom) Landau level.
   Full circles, open circles, diamonds, full squares, and open squares 
   mark data for angular momenta $L=0$, 3, 8, 12, and 18, respectively 
   (only selected values of $L$ are shown).}
\label{fig15}
\end{figure}

A closer inspection of figure~\ref{fig14} reveals a general tendency for 
the energy to increase with increasing $L$, which does not show up in 
the ${\cal G}(2p+1)$ spectrum.
The ${\cal G}(2p+1)$ spectrum predicts very well relative positions 
of energy levels with neighboring $L$'s, but, on the average, energy 
increases more quickly than ${\cal G}(2p+1)$ when $L$ is increased.
This is clearly visible in figure~\ref{fig15}(a), which shows energy 
of six electrons at $l={11\over2}$ ($\nu={2\over5}$) and ${15\over2}$ 
($\nu={1\over3}$) as a function of ${\cal G}(1)$.
States with different angular momenta $L$ are marked with different
symbols, and only five values, $L=0$, 3, 8, 12, and 18 are shown for 
clarity.
In the lowest LL, energy and ${\cal G}(1)$ and are quite well correlated 
within each $L$ subspace, and the relation between the two is almost 
identical for close values of $L$ (e.g., $L=0$ and 3).
However, for very different values of $L$ (e.g., $L=0$, 8, 12, and 18), 
the dependence of ${\cal G}(1)$ on energy changes considerably.
As found in figure~\ref{fig14}, in a pair of states with equal values of 
${\cal G}(1)$, the state with higher $L$ tends to have higher energy.
Clearly, this is due to the contributions of lower order terms in 
equation~\ref{eq7}.
It will be apparent from equation~\ref{eq99b} that the second highest 
term, ${\cal G}(2p+3)\,V(2p+3)$, increases with $L$ like, roughly, 
${\cal G}(2p+3)\sim L(L+1)$.

The similarity of the energy and ${\cal G}(2p+1)$ spectra makes it clear 
that a model pseudopotential with only one nonvanishing pseudopotential 
parameter, $V(1)>0$, reproduces the main features of the spectrum for
$\nu\ge{1\over3}$.
Similarly, the spectrum of a model pseudopotential with a hard core, 
$V(1)=\infty$, one finite parameter, $V(3)>0$, and all higher parameters
vanishing, resembles the low energy band of the Coulomb spectrum for 
${1\over3}\ge\nu\ge{1\over5}$.
In general, for the filling factor $\nu$ in the range $(2p+3)^{-1}<\nu
\le(2p+1)^{-1}$, the finite energy eigenstates of the hard core 
pseudopotential defined as
\begin{eqnarray}
   V^{(p)}_{\rm HC}({\cal R}<2p+1)&=&\infty,
\nonumber\\
   V^{(p)}_{\rm HC}({\cal R}=2p+1)&=&1,
\nonumber\\
   V^{(p)}_{\rm HC}({\cal R}>2p+1)&=&0
\end{eqnarray}
are very close to those of the Coulomb pseudopotential.
The dependence of finite eigenenergies of $V^{(p)}_{\rm HC}$ on angular 
momentum $L$ reproduces main features of the lowest band of the Coulomb 
spectrum.

Due to different behavior of the pseudopotential, the above conclusion 
does not generally hold for higher LL's.
The correlation between energy and ${\cal G}(1)$ for the same filling 
factors $\nu={2\over5}$ and ${1\over3}$ within the first excited LL 
($n=1$), plotted in figure~\ref{fig15}(c,d), is much worse than that for 
$n=0$ in figure~\ref{fig15}(a,b).
In particular, the lowest energy $L=0$ state is no longer the state with
the smallest ${\cal G}(1)$ at either filling.
Also, the Coulomb eigenstates in figure~\ref{fig15}(c,d) are not similar 
to those of a hard core repulsion.
For example, there is no Laughlin like state at $\nu={1\over3}$ with 
${\cal R}\ge3$ (instead, ${\cal G}(1)>0.06$ for all states) and no Jain 
like state at $\nu={2\over5}$ with ${\cal G}(1)\approx0.12$ (instead, 
${\cal G}(1)>0.19$ for all states).

As shown in figure~\ref{fig15}(e,f), the correlation between energy and 
${\cal G}(1)$ reappears in the second excited LL (Haldane and Rezayi 
1985a).
However, it is reversed and the low energy states have high values of 
${\cal G}(1)$.
At $\nu={2\over5}$, the Jain like state with ${\cal G}(1)\approx0.12$, 
maximally avoiding pair states with the smallest average separation and 
largest repulsion, is the highest energy state in its $L=0$ subspace.
Similarly, the highest $L=0$ state at $\nu={1\over3}$ is the Laughlin like 
state with ${\cal G}(1)\approx0.02$.

\begin{figure}[t]
\epsfxsize=3.4in
\epsffile{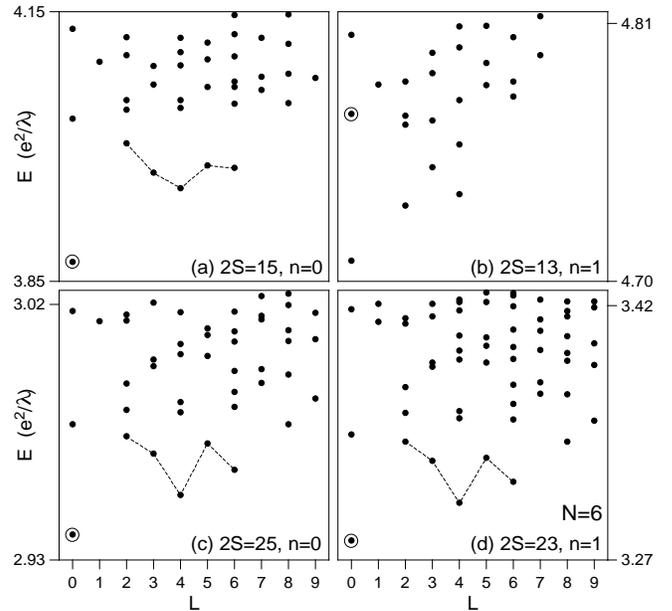}
\caption{
   The energy spectra of six electrons each with angular momentum
   $l={15\over2}$ (top) and $l={25\over2}$ (bottom), in the lowest
   (left) and first excited (right) Landau level.
   Open circles: states maximally avoiding pairs with largest repulsion.
   Dashed lines: states with one quasielectron--quasihole pair.}
\label{fig16}
\end{figure}

\begin{figure}[t]
\epsfxsize=3.4in
\epsffile{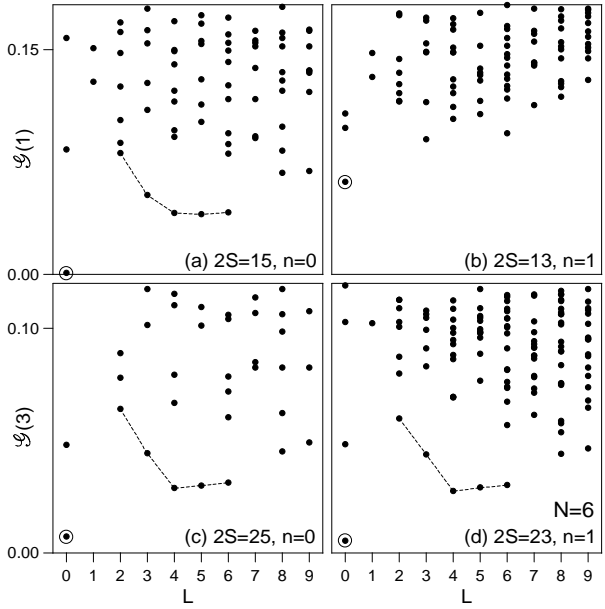}
\caption{
   The grandparentage coefficients ${\cal G}(1)$ and ${\cal G}(3)$ 
   for the eigenstates of six electrons each with angular momentum
   $l={15\over2}$ (top) and $l={25\over2}$ (bottom), in the lowest
   (left) and first excited (right) Landau level.
   Open circles: states maximally avoiding pairs with largest repulsion.
   Dashed lines: states with one quasielectron--quasihole pair.}
\label{fig17}
\end{figure}

The approximation of the Coulomb pseudopotential by the hard core 
pseudopotential, which gives almost exact many body eigenstates in the 
lowest LL and predicts the sequence of the Laughlin incompressible 
ground states, becomes valid in higher LL's at lower density (filling 
factor).
For $n=1$ and at fillings $\nu\le{1\over5}$, the second lowest band 
(${\cal R}\ge3$) couples to the next higher one (${\cal R}\ge1$). 
Interband coupling means here that the actual eigenstates are linear 
combinations of hard core eigenstates from both bands and the eigenstates 
originating from the ${\cal R}\ge3$ band of the hard core spectrum 
have some grandparentage from the ${\cal R}=1$ pair state.
However, as seen in figure~\ref{fig8} for only one (ground) state,
the band originating from the ${\cal R}\ge5$ band is (to a good 
approximation) uncoupled, i.e.\ its eigenstates indeed all have 
${\cal R}\ge5$ and are very close to the corresponding hard core states.
This occurs because the decoupling of the lowest band from the rest 
of the spectrum depends on the behavior of the pseudopotential $V$ 
at ${\cal R}\ge3$, where $V$ of $n=1$ is similar to that of $n=0$ 
(see figure~\ref{fig5}).
Figure~\ref{fig16} shows the energy spectra of six electrons each 
with $l={15\over2}$ (filling factor $\nu={1\over3}$) and $l={25\over2}$ 
($\nu={1\over5}$), for the lowest and first excited LL.
Figure~\ref{fig17} shows the corresponding spectra of ${\cal G}(1)$ 
and ${\cal G}(3)$.
States marked with open circles are states with the lowest ${\cal G}(1)$
for $l={15\over2}$ in frames (a) and (b) and states with ${\cal G}(1)
\approx0$ and the lowest ${\cal G}(3)$ for $l={25\over2}$ in frames (c) 
and (d).
Dashed lines connect the states that contain a single QP pair in the
mean field CF picture.
Clearly, even though the ground states in frames (a) and (b) both have 
$L=0$, the two spectra for $l={15\over2}$ are different.
For $n=1$, the band of states with one QP pair is absent, the ground 
state is not the one with lowest ${\cal G}(1)$, and none of the states 
has ${\cal G}(1)\approx0$.
On the other hand, the two spectra at $l={25\over2}$ in frames (c) and (d) 
are very similar.
Both contain the band of states with one QP pair, and have the Laughlin 
$\nu={1\over5}$ ground states with ${\cal G}(1)\approx{\cal G}(3)\approx0$.

\section{Relation between pseudopotential 
         and occurrence of incompressible ground states}
\label{secVII}

\subsection{Total angular momentum vs.\ average pair angular momentum}
\label{secVIIa}

A very useful operator identity
\begin{equation}
   \sum_{i<j} \hat{L}_{ij}^2 = \hat{L}^2 + N(N-2)\,\hat{l}^2
\label{eqthr}
\end{equation}
is straightforward to prove (W\'ojs and Quinn 1999a).
Here $\hat{L}=\sum_i\hat{l}_i$ and $\hat{L}_{ij}=\hat{l}_i+\hat{l}_j$.
Taking the expectation value of equation~\ref{eqthr} in the state 
$\left|l^N,L\alpha\right>$ gives
\begin{equation}
   \left<\right.\sum_{i<j}\hat{L}_{ij}^2\left.\right>
   =L(L+1)+N(N-2)\,l(l+1),
\label{threxp}
\end{equation}
which is independent of which multiplet $\alpha$ of a given angular 
momentum $L$ is being considered.
From equation~\ref{eq3} we also have
\begin{equation}
   \left<\right.\sum_{i<j}\hat{L}_{ij}^2\left.\right>
   ={N(N-1)\over2}\sum_{L_{12}}{\cal G}_{L\alpha}(L_{12})\,
   L_{12}(L_{12}+1).
\label{eq51}
\end{equation}
Combining the above two equations, a nontrivial condition on the 
allowed values of CFGP's is obtained.
Adding the normalization condition following from equation~\ref{eq6}, 
we have the following pair of constraints on the allowed CFGP's 
profiles ${\cal G}_{L\alpha}({\cal R})$ in a multiplet of a given $L$
\begin{eqnarray}
   \sum_{L_{12}}{\cal G}_{L\alpha}(L_{12})&&\,L_{12}(L_{12}+1)
\nonumber\\
   &&={L(L+1)+N(N-2)\,l(l+1)\over N(N-1)/2},
\label{eq99a}
\\
   \sum_{L_{12}}{\cal G}_{L\alpha}(L_{12})&&=1.
\label{eq99b}
\end{eqnarray}
The minimization of the total interaction energy in a Hilbert space
of a given $N$, $l$, $M$, and $L$ occurs through the optimization of 
the CFGP profile ${\cal G}({\cal R})$ (i.e., the pair correlation 
function), and must conform to the above constraints.

\subsection{Harmonic repulsive interaction}
\label{secVIIb}

It follows from equations~\ref{eq7}, \ref{threxp}, and \ref{eq51}, that 
if the pseudopotential were given by 
\begin{equation}
   V_{\rm H}(L_{12})=c_1+c_2\,L_{12}(L_{12}+1),
\end{equation}
all different multiplets with the same value of total angular momentum 
$L$ would be degenerate at the energy 
\begin{eqnarray}
   E_{L\alpha}&=&c_1N(N-1)/2
\nonumber\\
                      &+&c_2[L(L+1)+N(N-2)\,l(l+1)].
\end{eqnarray}
What is the physical meaning of the pseudopotential $V_{\rm H}$ which 
is linear in $\hat{L}_{12}^2$?
From equation~\ref{eqharm}, $V_{\rm H}$ is the harmonic interaction,
\begin{equation}
   V_{\rm H}(|{\bf r}_i-{\bf r}_j|)
   =c_1'-c_2'{|{\bf r}_i-{\bf r}_j|^2\over R^2},
\end{equation}
and, using equation~\ref{eqthr}, the many body harmonic interaction 
Hamiltonian can be written as
\begin{equation}
   H_{\rm H}=c_1N(N-1)/2+c_2\,N(N-2)\,l(l+1)+B\,\hat{L}^2,
\end{equation}
i.e.\ for the harmonic repulsive interaction within an isolated LL, 
each $L$ subspace is degenerate and the energy increases linearly 
with increasing $L(L+1)$.

The difference between the harmonic and actual pseudopotentials, 
the `anharmonic' contribution $V_{\rm AH}=V-V_{\rm H}$, lifts this 
degeneracy and the actual values of $E_{L\alpha}$ depend on how 
the values of ${\cal G}_{L\alpha}(L_{12})$ are distributed, not just 
on the average value of $\hat{L}_{12}^2$ for that value of $L$.
However, if the anharmonic correction $V_{\rm AH}$ is small, the 
ground state will have the lowest available value of angular momentum, 
$L=L^{\rm MIN}$.
If $V_{\rm AH}$ is not small, different multiplets with the same $L$ 
repel one another, and the splittings caused by $V_{\rm AH}$ can become 
large when $N_L$, the number of times the multiplet $L$ occurs, is large.
As a result, the lowest multiplet with certain angular momentum $L$ can 
have lower energy than multiplets of a smaller neighboring $L'$, for 
which $N_{L'}\ll N_L$.
In this case, a state with $L$ larger than $L^{\rm MIN}$ can become
the ground state.
For example, for the system of eight electrons at $2S=22$, the lowest 
energy multiplet at $L=4$ has lower energy than the multiplets at $L=0$, 
1, 2, and 3 (see figure~\ref{fig1}(b) and table~\ref{tab6}).
Even if $V_{\rm AH}$ is not small, if only $V(L_{12})$ increases with 
increasing $L_{12}$, then states with low angular momentum $L$ (and thus 
low average pair angular momentum $L_{12}$) will tend to have low energy, 
and states with high $L$ will tend to have high energy.

\begin{figure}[t]
\epsfxsize=3.4in
\epsffile{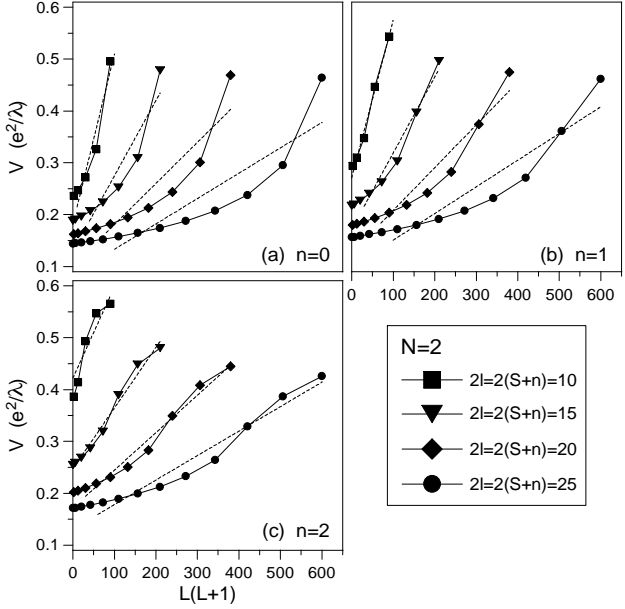}
\caption{
   The pseudopotentials of the Coulomb interaction in the lowest (a), 
   first excited (b), and second excited (c) Landau level, as a function 
   of squared pair angular momentum. 
   Squares: $l=5$, triangles: $l={15\over2}$, 
   diamonds: $l=10$, circles: $l={25\over2}$.
   The dashed lines: pseudopotentials corresponding to the the best 
   harmonic interaction fit of the six electron spectrum, as shown 
   by the dashed lines in figure~\protect\ref{fig19} for $l=5$.}
\label{fig18}
\end{figure}

How close is the actual Coulomb pseudopotential to the harmonic one?
The plots of $V$ given as a function of squared pair angular momentum 
$L(L+1)$, are shown in figure~\ref{fig18}.
The pseudopotentials for $n=0$ increase more quickly than linearly 
with increasing $L(L+1)$ in the entire range of $L$.
For $n=1$, they do so at low values of $L$, and the dependence
is almost linear close to ${\cal R}=1$.
And for $n=2$, $V$ becomes a sublinear function of $L(L+1)$ at high 
energy.
The dashed lines give the pseudopotentials of a harmonic interaction 
which correspond to the best fit to the six electron spectra.

\begin{figure}[t]
\epsfxsize=3.4in
\epsffile{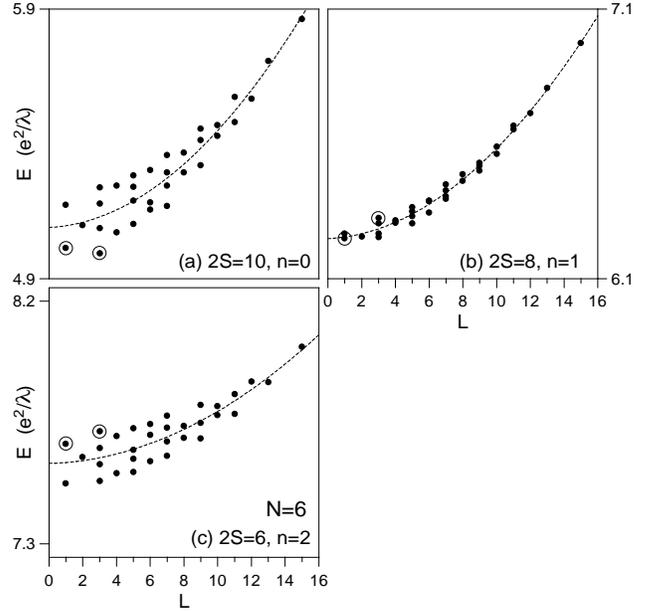}
\caption{
   The energy spectra of six electrons, each with $l=5$, in the 
   lowest (a), first excited (b), and second excited (c) Landau level.
   The best harmonic interaction fits to the Coulomb spectra (dots) 
   are marked with dashed lines (corresponding harmonic interaction 
   pseudopotentials are marked in figure~\protect\ref{fig18}).
   Open circles: states maximally avoiding pair states with largest 
   repulsion.}
\label{fig19}
\end{figure}

Examples of energy spectra of the six electron system in the lowest
($n=0$) and two excited ($n=1$ and 2) LL's approximated by the harmonic 
interaction are shown in figure~\ref{fig19} for $l=5$.
The general trend for the energy to increase with $L$ as well as the 
effects due to level repulsion caused by the anharmonicity of the 
pseudopotentials are visible.
In all frames, the highest energy state is the one with the highest 
$L$, and the lowest energy states have low $L$.
The spectrum for $n=1$ is less distorted from its harmonic fit than 
the spectra for $n=0$ and 2.
This reflects the fact that the corresponding pseudopotential, marked 
with squares in figure~\ref{fig18}(b), is closer to a harmonic one 
than the other two, also marked with squares in figure~\ref{fig18}(a) 
and (c).
For $n=1$ and 2, the ground state has the lowest available angular 
momentum $L=L^{\rm MIN}=1$.
For $n=0$, the anharmonicity of the pseudopotential is sufficiently 
large for the state with $L=3>L^{\rm MIN}$, to become the ground state 
due to the level repulsion ($N_3=4$ is larger than $N_1=2$ or $N_2=1$). 
Open circles in figure~\ref{fig19} mark the two states at $L=1$ and 3, 
which have the lowest ${\cal G}(1)$ of all states in the spectrum. 
For $n=0$ these states are predicted by the mean field CF picture as 
the states of two QE's in the $\nu={2\over5}$ state.

\subsection{Comparison with atomic system: Hund's rule}
\label{secVIIc}

\begin{figure}[t]
\epsfxsize=3.4in
\epsffile{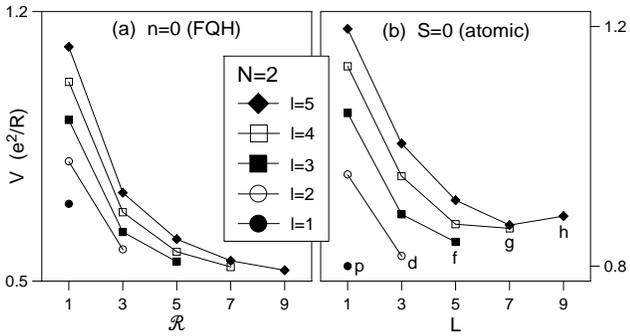}
\caption{
   The pseudopotentials $V$ of the Coulomb interaction potential 
   $V(r)=e^2/r$ for a pair of electrons each with angular momentum 
   $l$:
   (a) lowest Landau level, monopole harmonics, $n=0$ and $l=S$, 
   $V$ plotted as a function of relative pair angular momentum 
   ${\cal R}$;
   (b) atomic shell, spherical harmonics, $S=0$ and $l=n$, calculated 
   for a radial wavefunction which localizes electrons at radius $R$, 
   $V$ plotted as a function of pair angular momentum $L$.}
\label{fig20}
\end{figure}

The problem of electrons in a high magnetic field, occupying single 
particle states of the $n$th LL (monopole harmonics with $2S>0$, shell 
index $n\ll S$ and angular momentum $l=S+n$), can be compared to that 
of electrons in an atomic shell, occupying atomic states (spherical 
harmonics with $S=0$ and $l=n$).
In both cases the problem is that of $N$ electrons each with angular
momentum $l$ in a degenerate shell of states with different values of $m$.
However, the pseudopotential $V({\cal R})$ behaves very differently in
the two systems.
The comparison between the extreme $n=0$ and $S=0$ cases is presented 
in figure~\ref{fig20}.
The pseudopotentials for the lowest LL shell $V_{n=0}$ and for the atomic 
shell $V_{S=0}$, calculated for the same $l=S+n$, look quite similar when 
$V_{S=0}$ is plotted as a function of pair angular momentum $L$, and 
$V_{n=0}$ is plotted as a function of relative pair angular momentum 
${\cal R}=2l-L$.
Therefore, while $V_{n=0}$ decreases quickly with increasing ${\cal R}$
and attains the highest value at ${\cal R}=1$, the $V_{S=0}$ does just 
the opposite.

\begin{figure}[t]
\epsfxsize=3.4in
\epsffile{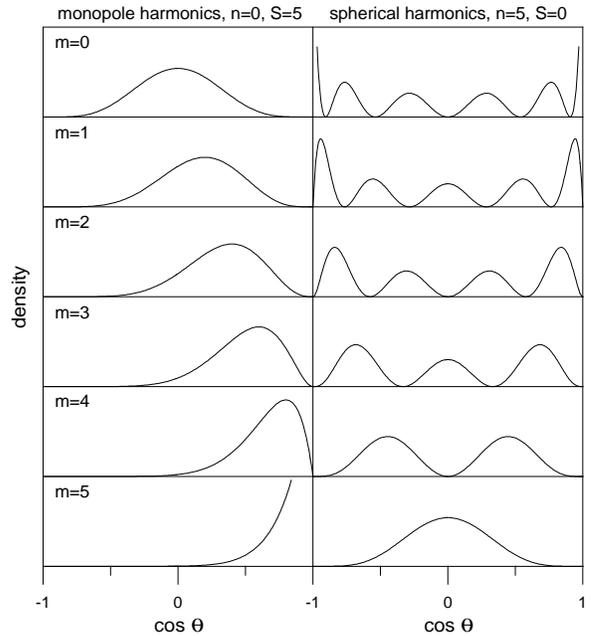}
\caption{
   The single particle density profiles:
   (a) lowest Landau level, monopole harmonics, $n=0$, $l=S$;
   (b) atomic shell, spherical harmonics, $S=0$, $l=n$,
   calculated for a radial wavefunction which localizes the electron at 
   radius $R$.}
\label{fig21}
\end{figure}

The pseudopotentials in both frames in figure~\ref{fig20} describe the 
same, Coulomb electron--electron interaction $V(r)=e^2/r$, and the 
origin of this difference lies in the very different Hilbert spaces.
The density profiles $\varrho_m(\cos\theta)$ for the single particle 
states in both cases are shown in figure~\ref{fig21}.
The $\theta$ is the standard spherical coordinate; $z=R\cos\theta$.
The average value of $z$ is (W\'ojs and Quinn 1998a)
\begin{equation}
   \left<S,l,m|z|S,l,m\right>={mS\over l(l+1)}R,
\end{equation}
and $\varrho_{-m}(z)=\varrho_m(-z)$ for the monopole harmonics, and 
$\varrho_{-m}(z)=\varrho_m(z)$ for the spherical harmonics.

The two electron state $\left|L,M\right>$ with maximum pair angular 
momentum $L=2l-1$ and $M=L$ is the single particle configuration 
$\left|m_1=l,m_2=l-1\right>$.
For the monopole harmonics, it has high Coulomb energy, as it corresponds 
to two electrons tightly packed around the north pole of the sphere.
On the other hand, the two electron state with minimum pair angular 
momentum, $L=0$, can be written as $\left|L=0,M=0\right>=\sum_m\zeta_m
\left|m_1=m,m_2=-m\right>$, i.e.\ in each contributing single particle 
configuration $\left|m_1,m_2\right>$ the two electrons have opposite 
$m$'s.
Opposite $m$'s mean opposite $\left<z\right>$'s and large spatial 
separation, and therefore the pair state with $L=0$ must have low 
interaction energy.

For the spherical harmonics, a similar analysis gives opposite answers.
The state $\left|m_1=l,m_2=l-1\right>$ with maximum allowed $L$ 
corresponds to two electrons spread over a large part of the sphere 
and avoiding one another (high density for $m=l$ occurs at $z$ 
corresponding to low density for $m=l-1$, and vice versa).
Therefore this state must have low Coulomb energy.
In the state with minimum $L=0$, built of single particle configurations 
$\left|m_1=m,m_2=-m\right>$, opposite $m$'s mean equal density profiles 
$\varrho(\cos\theta)$, and thus small average separation and high 
interaction energy.

In the case of an atomic system, the reasoning based on 
equation~\ref{eqthr} and the pseudopotential profile leads to the 
Hund's rule.
The multiplets with larger total angular momentum $L$ have, on the 
average, larger pair angular momenta $L_{ij}$ and thus lower energy.
There is only one multiplet with the maximum allowed total angular 
momentum $L=L^{\rm MAX}=Nl-N(N-1)/2$; it is a single particle 
`compact droplet' (maximum density) configuration, for $M=L$ equal 
to $\left|m_1,m_2,\dots,m_N\right>=\left|l,l-1,\dots,l-N+1\right>$.
It has the highest value of the average pair angular momentum and 
hence it is very likely to be the ground state.
A transition to a ground state at a neighboring lower $L$ would 
require strong anharmonicity of the pseudopotential.
Since relatively low multiplicities $N_L$ occur at $L$'s close to 
$L^{\rm MAX}$ ($N_{L^{\rm MAX}-1}=0$, $N_{L^{\rm MAX}-2}\le1$, 
$N_{L^{\rm MAX}-3}\le1$, etc.), $V_{\rm AH}$ does not affect the 
ordering of the levels at high $L$.
Despite this strong indication that the state with the largest $L$ 
has the lowest energy in atomic systems, Hund's rule is considered 
an empirical rule, that can be rigorously justified only by detailed 
numerical calculations.
It is also noteworthy that the atomic Hund's rule is usually of interest
only for rather low values of $l$ (up to the atomic $g$ or $h$ shell).

By analogy, the opposite rule can be formulated for monopole harmonics
(FQH system on a Haldane sphere), saying that the state with the maximum
$L$ has the highest energy.
Since for monopole harmonics the low energy states have low values 
of angular momentum $L$ (with large multiplicities $N_L$), the direct 
analog of the atomic Hund's rule (selecting the ground state) requires 
that correction $V_{\rm AH}$ is negligible.
Under this assumption it states that the state with lowest available $L$ 
has the lowest energy.
Both rules have been verified numerically.

For the Coulomb interaction acting in the space of monopole harmonics 
in the lowest LL, the assumption that $V_{\rm AH}$ is negligible does 
not hold and the multiplicities $N_L$ at low $L$ play a crucial role 
in determining low energy $L$ multiplets.
In such a general case, knowing which multiplet is the ground state or 
which multiplets form the low energy sector without performing detailed 
numerical calculations is a considerably more difficult task.
The prescription that the low energy states are found at those of low 
values of $L$ which correspond to large $N_L$ can be thought of as 
a more appropriate analog to the atomic Hund's rule.
As is the case with the atomic Hund's rule, it is an empirical rule 
that must be verified numerically.

Importantly, the $L$ multiplets for which $N_L$ is relatively large tend 
to reoccur at the same values of angular momentum $L$ when $2S$ is replaced
by $2S^*=2S-2p(N-1)$.
In table~\ref{tab6} we present, as an example, $N_L$ as a function of $L$ 
and $2S$ for a system of eight electrons.
\begin{table}
\caption{
   The number $N_L$ of independent multiplets at angular momentum $L$ 
   for eight electrons as a function of $2S$ for $0\le2S\le22$.
   Only values of $L$ up to 8 are included in the table.}
\begin{tabular}{r|rrrrrrrrr}
   $_{2S}\mbox{}^{L}$
   &0&1&2&3&4&5&6&7&8\\\hline
   0&&&\underline{1}&&&&&&\\
   1&\underline{1}&&\underline{1}&&\underline{1}&&&&\\
   2&\underline{1}&&&&&&&&\\
   3&\underline{1}&&\underline{1}&&\underline{1}&&&&\\
   4&\underline{1}&&\underline{1}&\underline{1}&\underline{1}&&
      \underline{1}&&\\
   5&\underline{1}&&\underline{1}&&\underline{1}&&\underline{1}&&\\
   6&&&&&\underline{1}&&&&\\
   7&\underline{1}&&&&&&&&\\
   8&&&&&\underline{1}&&&&\\
   9&\underline{1}&&\underline{1}&&\underline{1}&&\underline{1}&&1\\
   10&\underline{1}&&\underline{1}&\underline{1}&\underline{2}&1&
      \underline{2}&1&1\\
   11&\underline{2}&&\underline{3}&1&\underline{4}&2&4&2&4\\
   12&\underline{2}&1&4&3&6&5&7&5&7\\
   13&\underline{4}&1&\underline{7}&5&\underline{11}&7&13&9&13\\
   14&4&3&\underline{10}&9&16&14&19&17&21\\
   15&\underline{7}&4&\underline{16}&13&\underline{25}&21&31&26&35\\
   16&\underline{8}&8&21&22&35&33&45&42&51\\
   17&\underline{12}&10&\underline{32}&30&\underline{51}&48&66&61&77\\
   18&\underline{13}&17&\underline{42}&\underline{45}&\underline{69}&
      70&\underline{91}&90&108\\
   19&\underline{20}&22&\underline{58}&61&\underline{96}&95&\underline{128}
      &124&152\\
   20&22&33&75&85&\underline{126}&133&169&173&205\\
   21&\underline{31}&42&101&111&168&175&227&230&277\\
   22&36&59&126&150&\underline{215}&233&294&307&360
\end{tabular}
\label{tab6}
\end{table}
The values of $2S$ go from zero to twenty two; the values of $L$ are shown
up to eight.
The $L$ spaces which are predicted by the CF picture to form the lowest
energy band are underlined.
Clearly, they coincide with relatively high multiplicities $N_L$ at the
lower values of $L$.
Notice, for example, that the high $N_L$ values at $2S=19$, 20, and 21 
appear at the same angular momenta $L$ as the allowed multiplets at 
$2S^*=5$, 6, and 7, respectively.

\subsection{Connection between Hund's rule
            and avoiding pair states of large repulsion}
\label{secVIId}

What is the connection between the two predictions of low energy states
discussed earlier, (i) the Hund's rule argument selecting multiplets at 
low $L$ with high $N_L$ and (ii) the argument selecting multiplets that 
avoid large fractional grandparentage from pair states with largest 
repulsion? 
Let us first notice that whether a many body state without grandparentage 
from certain pair states belongs to the Hilbert space of given $N$, $l$, 
and $L$ depends critically on $N_L$.
It follows from equations~\ref{eq3} and \ref{eq52} that a multiplet 
with ${\cal G}({\cal R})=0$ (e.g., for ${\cal R}=1$) can be constructed 
if the degeneracy $N_L$ exceeds $N_{\cal R}$, the number of terms 
$(L',\alpha')$ in equation~\ref{eq3} with $L_{12}$ corresponding to 
${\cal R}$.
For example, for $L=0$, the addition of angular momentum vectors, 
${\bf L}={\bf L}_{12}+{\bf L}'$, selects only one value of $L'$ equal 
to $L_{12}$.
In this case, it is guaranteed that $N_{\cal R}$ does not exceed 
$N'_{L_{12}}$, the number of all $L'=L_{12}$ multiplets of $N-2$ 
electrons each with angular momentum $l$.
The actual value of $N_{\cal R}$ can be smaller than $N'_{L_{12}}$ 
because of the Pauli exclusion principle, which eliminates some of 
the combinations of ${\bf L}'$ and ${\bf L}_{12}$.
However, $N_{L=0}>N'_{L_{12}}$ guarantees that a multiplet $\left|l^N,
0\alpha\right>$, a linear combination of terms in equation~\ref{eq3}, 
can be constructed, for which the coefficients $G_{0\alpha,L_{12}\alpha'}
({\cal R})$ vanish simultaneously for all $\alpha'$ and therefore so does 
the coefficient ${\cal G}_{0\alpha}({\cal R})$.

In general, it is difficult to determine $N_{\cal R}$ by adding 
dimensions of all relevant $L'$ spaces of $N-2$ electrons because 
of the Pauli principle which imposes additional constraints on 
CFGP's in equation~\ref{eq3}.
However, one can calculate the matrix ($\alpha$ vs.\ $L'\alpha'$) of 
coefficients $G_{L\alpha,L'\alpha'}({\cal R})$ for all multiplets of 
given $L$ (for any choice of basis states $\alpha$, not necessarily 
the interaction eigenstates), and determine $N_{\cal R}$ directly.
It is clear that $N_L$ must exceed certain mimimum value for the 
occurrence of $L$ multiplets which avoid grandparentage from certain 
(strongly repulsive) pair states.
It is also clear that the minimum $N_L$ that is required to exceed 
$N'_{L_{12}}$ increases with increasing $L$ since a larger number of 
angular momenta $L'$ satisfy the addition rule, $|L'-L_{12}|\le L\le 
L'+L_{12}$, for larger $L$.
If the multiplets with ${\cal R}\ge3$, 5, \dots\ can be constructed 
(belong to the Hilbert space of given $N$, $l$, and $L$), they will be 
the lowest energy eigenstates of the hard core interaction defined in 
equation~\ref{eqvhc}.
Hence, the above discussion explains the occurrence of such eigenstates 
at those of low values of $L$ which have high multiplicity $N_L$.

Another problem that still needs clarification is whether the multiplets 
with ${\cal R}\ge3$, 5, \dots\ are the eigenstates of the actual (not 
strictly hard core) interaction pseudopotential $V({\cal R})$ (e.g.\ 
the Coulomb interaction in a given LL), and if they have low energy.
In other words, what is the relevant measure of the `short range' 
character of electron--electron interaction in the lowest LL?
Or, what is the condition for $V({\cal R})$ to act like hard core 
repulsion and have the energy spectrum characteristic of the FQH effect, 
with low energy states that have ${\cal R}\ge3$, 5, \dots?
Clearly, whether the ground state and other low lying multiplets tend 
to avoid grandparentage from pair states with ${\cal R}=1$, 3, \dots\ 
depends not only on whether $V({\cal R})$ is a decreasing function of 
${\cal R}$, but on how quickly it decreases with ${\cal R}$ as well.
This is because the sequence of CFGP's of a given eigenstate $\left|L
\alpha\right>$ are mutually connected through the normalization condition 
given by equations~\ref{eq99a}, and the nontrivial condition~\ref{eq99b}.
For example, it turns out that the $\nu={1\over3}$ state with ${\cal G}(1)
\approx0$ always has the largest ${\cal G}(3)$ of all states.
Therefore, $V({\cal R})$ must decrease sufficiently quickly with 
increasing ${\cal R}$ for the state with ${\cal R}\ge3$ to be the 
ground state at the $\nu={1\over3}$ filling.

\subsection{Definition of short range pseudopotential}
\label{secVIIe}

\begin{figure}[t]
\epsfxsize=3.4in
\epsffile{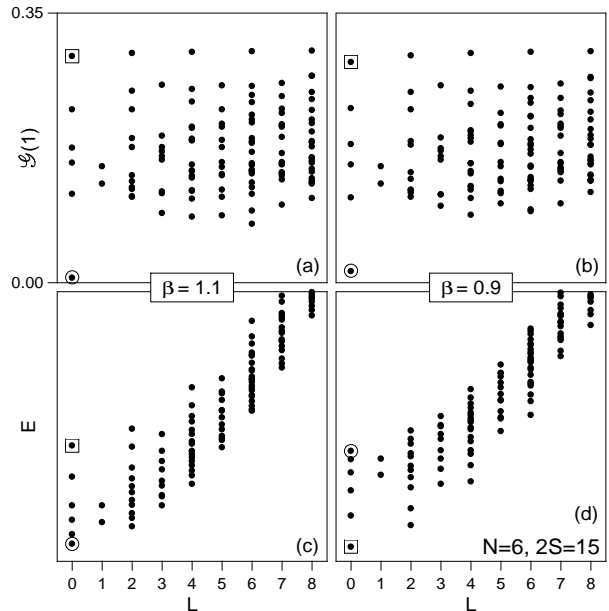}
\caption{
   The eigenenergies, (top), and the grandparentage coefficients 
   ${\cal G}(1)$, (bottom), for a system of six electrons each with 
   angular momentum $l={15\over2}$. 
   The interaction pseudopotential is 
   $V_\beta(L)=[L(L+1)]^\beta$, with $\beta=1.1$ (left) and 0.9 (right).
   The large open circles and squares mark the $L=0$ eigenstates 
   with minimum and maximum ${\cal G}(1)$, respectively.}
\label{fig22}
\end{figure}

The condition for the occurrence of the Laughlin incompressible 
$\nu=(2p+1)^{-1}$ ground states with ${\cal G}({\cal R}<2p+1)\approx0$ 
(and generally, for the occurrence of low energy states with 
${\cal G}({\cal R}<2p+1)\approx0$ and low ${\cal G}(2p+1)$ for 
$\nu<(2p+1)^{-1}$), is that pseudopotential $V(L)$ increases more quickly 
than linearly with increasing $L(L+1)$.
In the two top frames of figure~\ref{fig22} we show the energy spectra 
of a system of six electrons each with angular momentum $l={15\over2}$, 
calculated for a model pseudopotential 
\begin{equation}
   V_\beta(L)=[L(L+1)]^\beta
\end{equation}
with $\beta>1$ and $\beta<1$.
In the two bottom frames we plot the corresponding spectra of the CFGP 
corresponding to the highest pseudopotential parameter, ${\cal G}(1)$.
The ${\cal G}(1)$ spectra look quite similar for $\beta=1.1$ and 0.9.
In particular, in both cases there is one state in the spectrum 
(marked with a large open circle) whose ${\cal G}(1)$ almost vanishes.
At first sight, the energy spectra also look similar.
Both of them reveal overall tendency to increase energy with increasing 
$L$, and in both of them the larger width of $L$ subspectra coincides with 
larger $N_L$.
However, a closer inspection shows that the two spectra look like one 
another's vertical reflections.
For $\beta>1$, the states with low ${\cal G}(1)$ tend to have low energy.
For example, within the $L=0$ subspace, the state with ${\cal G}(1)
\approx0$ (large open circle, this is the Laughlin like $\nu={1\over3}$ state) 
has the lowest energy, and the state with the maximum ${\cal G}(1)\approx
0.3$ (large open square) has the highest energy.
On the contrary, for $\beta<1$, the states with low ${\cal G}(1)$ tend to 
have high energy.
For example, for $L=0$, the state with minimum ${\cal G}(1)$ has the 
highest energy and vice versa.
Clearly, the behavior of energy as a function of ${\cal G}(1)$ is opposite 
for $\beta>1$ and $\beta<1$.
This can be demonstrated even more clearly by comparing the expectation 
values of the $V_\beta$ interaction in the same states (instead of 
comparing the eigenspectra).
In this case the ordering of energies within each $L$ subspace is exactly 
reversed.

The exponent $\beta$ is the relevant measure of the `short range' 
character of a pseudopotential $V_\beta$.
The condition given by equation~\ref{eqvhc} that has been used to define 
an ideal short range (hard core) pseudopotential throughout this paper 
can be rewritten as $\beta\gg1$.
The pseudopotentials with $\beta>1$ define a class of `short range' 
repulsive interactions characterized by similar behavior of energy 
spectra and wavefunctions. 
For $\beta\rightarrow\infty$, the wavefunctions and structure of energy 
spectra converge to those of the model interaction in equation~\ref{eqvhc};
at the filling factor $\nu=(2p+1)^{-1}$ the ground state is given exactly 
by the Laughlin wavefunction (or by its spherical form given in 
(Haldane 1983)).
The pseudopotentials $V_\beta$ with $\beta<1$ belong to a separate
class of interactions, characterized by their own (common) behavior of 
energy spectra and wavefunctions (W\'ojs and Quinn 1998a), different from 
those of the short range class with $\beta>1$.
In particular, Laughlin incompressible $\nu=(2p+1)^{-1}$ ground states 
with ${\cal R}\ge2p+1$ occur only for $\beta>1$.
The harmonic interaction with $\beta=1$ separates those two classes and
does not belong to either one.

\subsection{Pseudopotentials of other 2D systems}
\label{secVIIf}

\begin{figure}[t]
\epsfxsize=3.4in
\epsffile{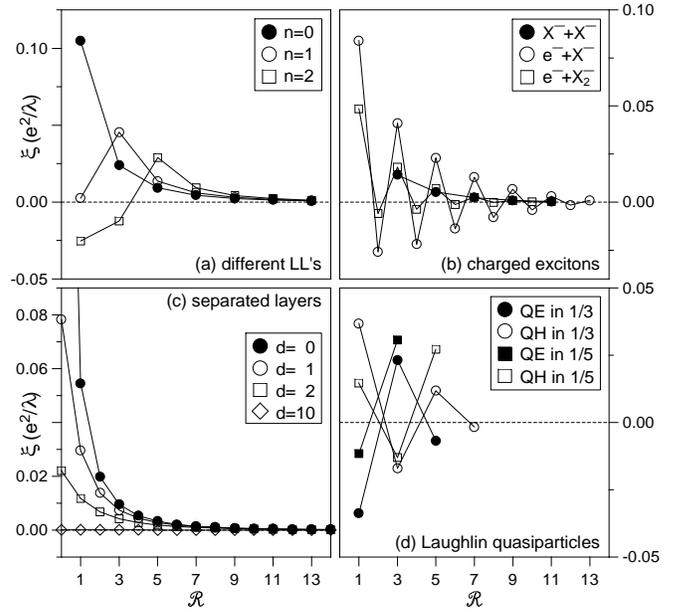}
\caption{
   The anharmonicity parameter $\xi$ as a function of relative 
   angular momentum ${\cal R}$ for pseudopotentials of different 
   electronic systems in a high magnetic field:
   (a) electrons in different Landau levels;
   (b) electrons and charged excitons in the lowest Landau level;
   (c) electrons in two parallel 2D layers separated by $d$ magnetic
   lengths;
   (d) Laughlin quasiparticles in the $\nu={1\over3}$ and ${1\over5}$
   ground states.}
\label{fig23}
\end{figure}

The Coulomb pseudopotential for the lowest LL is not strictly of the 
form $V_\beta(L)$.
However, as shown in figure~\ref{fig18}(a), it increases more quickly 
than linearly with an increase of $L(L+1)$ in entire range of $L$.
In consequence, the low energy states are those with ${\cal G}(1)\approx
{\cal G}(3)\approx\dots\approx{\cal G}(2p-1)\approx0$ and the lowest 
value of ${\cal G}(2p+1)$, and the $L=0$ ground states at $2S=(2p+1)(N-1)$ 
are Laughlin incompressible $\nu=(2p+1)^{-1}$ states.
In general, the low lying states of an interacting many body system at 
filling factor $\nu\sim(2p+1)^{-1}$ tend to have Laughlin correlations 
(the states with lowest energy have vanishing grandparentage from pair 
states with ${\cal R}<2p-1$ and smallest grandparentage from ${\cal R}=
2p-1$), if the pseudopotential $V({\cal R})$ decreases as a function 
of ${\cal R}$ in the entire range, and decreases more quickly than the 
harmonic pseudopotential $V_{\rm H}$ in the vicinity of ${\cal R}=2p+1$.
On a sphere, $V_{\rm H}$ increases linearly as a function of the squared 
pair angular momentum $L(L+1)$; on a plane it decreases linearly as 
a function of the angular momentum of the relative motion.
The condition for Laughlin correlations can be conveniently expressed 
in terms of the following anharmonicity parameter
\begin{equation}
   \xi({\cal R})=V({\cal R})-V_{\rm H}({\cal R}), 
\end{equation}
where $V_{\rm H}({\cal R})$ is the harmonic extrapolation of $V({\cal R}+4)$ 
and $V({\cal R}+2)$ at ${\cal R}$.
The condition states that Laughlin correlations (avoiding pairs with 
${\cal R}\le2p-1$) occur at $\nu\sim(2p+1)^{-1}$ if $\xi(2p-1)>0$.
In figure~\ref{fig23} we plot $\xi({\cal R})$ for a number of different 
2D electron systems in a high magnetic field. 
By analogy to the electron gas in the lowest LL, one could expect Laughlin 
like correlations in these systems, and we try to interpret them in terms 
of mean field CF's.

As shown in figure~\ref{fig23}(a) and in figure~\ref{fig18}(b,c), the 
electron pseudopotential in excited LL's is of the short range type
only for ${\cal R}\ge2n+1$.
In consequence, the ground states at Laughlin--Jain filling factors 
$\nu\ge(2n+1)^{-1}$ do not have Laughlin correlations (in contrast 
to the states at the same filling of the lowest LL).
Only at lower filling factors, $\nu<(2n+1)^{-1}$, where the part of 
the pseudopotential which does not decrease quickly enough with
increasing ${\cal R}$ is completely avoided and does not affect the 
lowest energy eigenstates, do these eigenstates have low grandparentage 
from pair states with large repulsion.
In particular, the Laughlin like incompressible ground states occur 
only at $\nu<(2n+1)^{-1}$.
This explains the compressibility of the $\nu=2+{1\over3}$, $4+{1\over3}$, 
and $4+{1\over5}$ ground states (or, at least, different correlations and 
thus different origin of their incompressibility), and the 
incompressibility of the $\nu=2+{1\over5}$ ground state, observed in 
figure~\ref{fig10}.

Another example of a system interacting through the short range 
pseudopotential is the system of charged excitons ($X^-$, two electrons 
bound to a valence hole) or biexcitons ($X_2^-$, three electrons bound 
to two valence holes) in the lowest LL, which has been shown to have 
Laughlin like incompressible ground states (W\'ojs, Hawrylak, and 
Quinn 1998c, W\'ojs, Hawrylak, and Quinn 1999b). 
This is confirmed in figure~\ref{fig23}(b), where we also plot 
$\xi({\cal R})$ calculated for the interaction of an $X^-$ or an $X_2^-$ 
with an electron ($e^-$).
Note that for a pair of distinguishable particles, ${\cal R}$ can take 
on any integer value, and that the pseudopotentials involving $X^-$ or 
$X_2^-$ have hard core ($V=\infty$) at a number of smallest values of 
${\cal R}$.
Clearly, the Laughlin like $e^-$--$X^-$ or $e^-$--$X_2^-$ correlations 
described by a Jastrow prefactor in the wavefunction will only occur at 
odd values of ${\cal R}$ (W\'ojs {\sl et al.}\ 1999c).

If electrons are confined in parallel 2D layers separated by a small
distance $d$, the inter-layer repulsion $V_d(r)=e^2/\sqrt{r^2+d^2}$ 
can result in the inter-layer Laughlin correlations, unless $d$ is 
larger than the characteristic separation between electrons in each 
layer ($\sim\sqrt{2\pi/\nu}\,\lambda$).
The plots of $\xi({\cal R})$ for the pseudopotentials $V_d({\cal R})$ 
in the lowest LL are shown in figure~\ref{fig23}(c).
When $d$ is large, $V_d(r)\approx(1-{1\over2}(r/d)^2+\dots)/d$ becomes 
essentially harmonic at small $r$, $V_d({\cal R})$ becomes harmonic 
at small ${\cal R}$ and the inter-layer correlations disappear.
Since $V_d(r)$ is a good approximation to an effective 2D potential
in a quasi-2D layer of finite width ($\sim5d$), figure~\ref{fig23}(c)
shows also the destruction of the FQH effect in a single wide quantum 
well (Shayegan {\sl et al.}\ 1990).

The CF hierarchy uses the mean field approach for the QP's and therefore 
should fail when applied to partially filled QP shells unless the QP 
pseudopotential has short range.
In states with completely filled QE shells (where $\nu_{\rm QE}$ is
an integer), the gap for creating a new type of QE--QH pair makes
the nondegenerate $L=0$ ground state an incompressible fluid state
regardless of the form of the QE pseudopotential.
For example, the Jain incompressible $\nu={2\over5}$ state is obtained 
when QE's of the $\nu={1\over3}$ parent state fill one shell ($\nu_{\rm 
QE}=1$).
For partially filled QP shells, the CF hierarchy correctly predicts 
daughter incompressible ground states only at certain fractional QP 
filling factors but not at others.
A quick look at the QP pseudopotentials in figure~\ref{fig11}(a,g) for 
ten electrons (as well as in figures~\ref{fig1}(c) and \ref{fig2}(d,h) 
for eight electrons) allows the prediction of filling factors at which 
the QP's indeed form a Laughlin ground state. 
In figure~\ref{fig23}(d) we plot $\xi({\cal R})$ for QP's of Laughlin 
$\nu={1\over3}$ and ${1\over5}$ states, obtained in diagonalization of 
eleven and eight electron systems, respectively.
It can be readily seen that Laughlin QH's should form a stable Laughlin
$\nu_{\rm QH}={1\over3}$ state of their own.
It follows from equation~\ref{eqjain} that the $\nu_{\rm QH}={1\over3}$ 
daughter state of the $\nu={1\over3}$ parent state corresponds to Jain 
$\nu={2\over7}$ state of electrons. 
Indeed, this state is an incompressible eight electron ground state 
in figure~\ref{fig1}(f).
On the other hand, the $\nu_{\rm QH}={1\over5}$ QH state and the 
corresponding $\nu={4\over13}$ electron state will be compressible.
Indeed, the eight electron ground state in figure~\ref{fig1}(d) does 
not even have $L=0$.
The $\nu_{\rm QH}={1\over7}$ might be incompressible but with a much 
smaller gap than that of $\nu_{\rm QH}={1\over3}$, what would lead to 
weak incompressibility of the $\nu={6\over19}$ electron state.
Indeed, the gap above the $L=0$ ground state of six electrons at 
$2S=17$ is very small.
For partially filled QE shells, the $\nu_{\rm QE}={1\over3}$ 
($\nu={4\over11}$) and $\nu_{\rm QE}={1\over7}$ ($\nu={8\over23}$) 
states are expected to be compressible, and the $\nu_{\rm QE}={1\over5}$ 
($\nu={6\over17}$) state could be weakly incompressible.
These predictions are in perfect agreement with numerical results 
for finite systems (W\'ojs and Quinn 2000), and we presume that taking 
into account the behavior of pseudopotentials of interaction between 
QE's and between QH's in different stable Laughlin states on all levels 
of hierarchy explains naturally all observed odd denominator FQH fillings 
and allows the prediction of their relative stability without using trial 
wavefunctions involving multiple LL's and projections onto the lowest LL.
The inconsistencies of the original QP hierarchy picture (Haldane 1983, 
Laughlin 1984, Halperin 1984): the appearance of some observed fractions 
on high hierarchy levels and the actual compressibility of some fractions 
predicted on lower levels, are removed by noticing that Laughlin QP's 
of a given type form incompressible Laughlin states of their own only 
at certain filling factors.

\subsection{Prescription for low energy multiplets}
\label{secVIIg}

The discussion presented in the preceding sections can be summarized 
in the form of a general prescription for the angular momentum multiplets 
forming the low energy sector in FQH systems.

(i)
The pseudopotential $V({\cal R})$ describing the Coulomb repulsion in 
an isolated (lowest or excited) LL decreases when relative pair angular 
momentum ${\cal R}$ increases, i.e.\ when the pair angular momentum 
$L_{12}$ decreases.

(ii)
Multiplets with lower total angular momentum $L$ have lower expectation 
value of the pair angular momentum $L_{12}$, and thus lower energy.

(iii)
The energy levels at the same $L$ repel one another due to the 
anharmonicity of $V({\cal R})$.
As a result, low values of total angular momentum $L$ for which many 
independent multiplets occur are more likely to have some states at 
lower energy than neighboring $L$ values with few multiplets.

(iv)
Relatively higher multiplicities $N_L$ tend to reoccur at the same 
values of $L$ for single particle angular momenta $l^*=l-p(N-1)$.
These values coincide with predictions of the mean field CF picture.

(v)
The many body Hilbert spaces corresponding to low angular momenta 
$L$ with large multiplicities $N_L$ (as predicted by the mean field 
CF picture) contain some states with small grandparentage from pair 
states of largest repulsion.

(vi)
If $V({\cal R})$ decreases more quickly with decreasing ${\cal R}$ than 
the harmonic pseudopotential, the low lying many body states avoid 
grandparentage from pair states of largest repulsion, and thus occur at 
total angular momenta predicted by the mean field CF picture.

(vii)
The gap above the low energy states that avoid grandparentage from pair 
states of largest repulsion is governed by the appropriate difference of 
pseudopotential parameters.
This gap does not collapse in the thermodynamic limit.

(viii)
At filling factors at which the low energy band separated from the rest 
of the spectrum by a gap contains only a nondegenerate (singlet) $L=0$ 
ground state, the system is incompressible and exhibits the FQH effect.

\section{Summary}
\label{secVIII}

We have shown that the success of the mean field composite Fermion 
(CF) picture in correctly and simply selecting the band of lowest 
energy multiplets of fractional quantum Hall (FQH) systems is not 
due to a cancellation between Coulomb and Chern--Simons interactions 
among fluctuations, which are described by totally different energy 
scales.
The reason for the success is related to the nature of the Coulomb 
pseudopotential $V({\cal R})$ in the lowest Landau level (LL).

We have identified an exact dynamical symmetry of the hard core 
repulsive (HCR) pseudopotential.
Due to this symmetry, the many body energy spectrum splits into bands
of eigenstates which avoid an increasing number of pseudopotential 
parameters of largest repulsion (the wavefunctions of these eigenstates 
contain Jastrow prefactors $\prod_{i<j}(z_i-z_j)^m$ with increasing 
exponents $m$).
The bands are separated by gaps which are associated with the difference 
of appropriate pseudopotential parameters and do not collapse in the 
thermodynamic limit.
The incompressibility of Laughlin $\nu=(2p+1)^{-1}$ states in a system
with HCR interactions results from the fact that the 
nondegenerate ($L=0$) ground state is the only state in its (lowest 
energy) band at these filling factors.
The mean field CF picture can be applied to such systems.

We have defined the class of `short range' (SR) pseudopotentials 
$V({\cal R})$, for which the Laughlin correlations (avoiding strongly 
repulsive pair states) minimize the total interaction energy.
The occurrence of distinct bands and Laughlin--Jain incompressible 
ground states in the energy spectrum of systems with SR interactions 
is a consequence of weakly broken dynamical symmetry of the HCR.
The pseudopotential $V({\cal R})$ has the SR character in a given 
range of relative pair angular momentum ${\cal R}$ if $V({\cal R})$ 
decreases in this range more quickly as a function of ${\cal R}$ than 
the harmonic pseudopotential.
The Coulomb repulsion in the lowest LL belongs to the SR class in 
entire range of ${\cal R}$, and hence Laughlin correlations occur 
at all Laughlin filling factors $\nu=(2p+1)^{-1}$.

We have found that the pseudopotentials in excited LL's decrease more 
slowly with increasing ${\cal R}$ and do not have SR character at the 
smallest values of ${\cal R}$.
As a result, the Laughlin correlations occur in excited LL's only 
below certain filling factor.
For example, we have shown that the $\nu=2+{1\over3}$ state does 
not have Laughlin correlations in the first excited LL, while 
$\nu=4+{1\over3}$ and $4+{1\over5}$ states do not have such correlations 
in the second excited LL.
On the other hand, the $\nu=2+{1\over5}$ state has Laughlin correlations 
and an excitation gap comparable to the $\nu={1\over5}$ state.
Because the mean field CF model describes systems with Laughlin 
correlations, it is only valid at lower fillings of excited LL's.

The CF hierarchy uses the mean field approach for the quasiparticles 
(QP's) and therefore should fail unless the QP pseudopotential has SR.
We have found that QP's have Laughlin correlations at some of the Laughlin 
filling factors but not at others.
This explains incompressibility of hierarchy ground states at 
$\nu={2\over7}$ and compressibility at such hierarchy fractions as 
$\nu={4\over11}$ or ${4\over13}$.
Also, since the Laughlin quasielectron (QE) and quasihole (QH) energies 
are governed by different electron pseudopotential parameters, the QE 
energy is higher than the QH energy.

We have also studied the validity of the atomic Hund's rule for 
systems with different pseudopotentials and shown that a modified Hund's 
rule remains valid for FQH systems on a Haldane sphere.
According to this rule, the FQH states with small total angular 
momentum $L$ tend to have lower energy than states with large $L$.
This rule is strict for the harmonic interaction for which energy
is completely independent of correlations.
Strong anharmonicity of the pseudopotential can invalidate this rule 
and favor either Laughlin correlated states at low $L$ with large 
number of multiplets if the pseudopotential has SR, or other 
type of correlations (e.g., possible pairing) if the pseudopotential 
is subharmonic.

\section*{Acknowledgment}

The authors gratefully acknowledge the support of Grant DE-FG02-97ER45657
from the Materials Science Program -- Basic Energy Sciences of the US
Department of Energy.
They wish to thank W. Bardyszewski, P. Hawrylak, D. C. Marinescu, 
P. Sitko, I. Szlufarska, and K.-S. Yi for helpful discussions on 
different aspects of this work.
A.W. acknowledges partial support from the Polish Sci. Comm. (KBN) Grant 
2P03B11118.

\section*{References}

\begin{raggedright}
Belkhir, L., and Jain, J., K., 1993,
{\sl Phys. Rev. Lett.}, {\bf70}, 643.
\\[0.5ex]
Chen, X. M., and Quinn, J. J., 1996,
{\sl Solid State Commun.}, {\bf92}, 865.
\\[0.5ex]
Cowan, R. D., 1981,
{\sl The Theory of Atomic Structure and Spectra} 
(Berkeley: University of California Press). 
\\[0.5ex]
Dirac, P. A. M., 1931,
{\sl Proc. R. Soc. London}, Ser. A {\bf133}, 60.
\\[0.5ex]
Fano, G., Ortolani, F., and Colombo, E., 1986,
{\sl Phys. Rev. B}, {\bf34}, 2670.
\\[0.5ex]
Haldane, F. D. M., 1983,
{\sl Phys. Rev. Lett.}, {\bf51}, 605.
\\[0.5ex]
Haldane, F. D. M., and Rezayi, E. H., 1985a,
{\sl Phys. Rev. Lett.}, {\bf54}, 237.
\\[0.5ex]
Haldane, F. D. M., and Rezayi, E. H., 1985b
{\sl Phys. Rev. B}, {\bf31}, 2529.
\\[0.5ex]
Haldane, F. D. M., 1987,
{\sl The Quantum Hall Effect}, 
edited by R. E. Prange and S. M. Girvin
(New York: Springer-Verlag), chapter 8, pp. 303--352.
\\[0.5ex]
Haldane, F. D. M., and Rezayi, E. H., 1988
{\sl Phys. Rev. Lett.}, {\bf60}, 956.
\\[0.5ex]
Halperin, B. I., 1983,
{\sl Helv. Phys. Acta} {\bf56}, 75.
\\[0.5ex]
Halperin, B. I., 1984,
{\sl Phys. Rev. Lett.}, {\bf52}, 1583.
\\[0.5ex]
Halperin, B. I., Lee, P. A., and Read, N., 1993,
{\sl Phys. Rev. B}, {\bf47}, 7312.
\\[0.5ex]
Haydock, R., 1980,
{\sl Solid State Physics}, {\bf35} 215.
\\[0.5ex]
He, S., Xie, X., and Zhang, F., 1992,
{\sl Phys. Rev. Lett.}, {\bf68}, 3460.
\\[0.5ex]
Jain, J., K., 1989,
{\sl Phys. Rev. Lett.}, {\bf63}, 199.
\\[0.5ex]
von Klitzing, K., Dorda, G., and Pepper, M., 1980,
{\sl Phys. Rev. Lett.}, {\bf45}, 494.
\\[0.5ex]
Lanczos, C., 1950,
{\sl J. Res. Natn. Bur. Stand.}, {\bf45}, 255.
\\[0.5ex]
Laughlin, R. B., 1983a,
{\sl Phys. Rev. Lett.}, {\bf50}, 1395.
\\[0.5ex]
Laughlin, R. B., 1983b,
{\sl Phys. Rev. B}, {\bf27}, 3383.
\\[0.5ex]
Laughlin, R. B., 1984,
{\sl Surf. Sci.}, {\bf142}, 163.
\\[0.5ex]
Lopez A., and Fradkin E., 1991,
{\sl Phys. Rev. B}, {\bf44}, 5246.
\\[0.5ex]
MacDonald A. H., and Girvin, S. M., 1986, 
{\sl Phys. Rev. B}, {\bf33}, 4009.
\\[0.5ex]
Moore, G., and Read, N., 1991,
{\sl Nucl. Phys. B}, {\bf360}, 362.
\\[0.5ex]
Morf, R., and Halperin, B. I., 1986,
{\sl Phys. Rev. B}, {\bf33}, 2221.
\\[0.5ex]
Rezayi, E. H., and MacDonald, A. H., 1991,
{\sl Phys. Rev. B}, {\bf44}, 8395.
\\[0.5ex]
de Shalit, A., and Talmi, I., 1963,
{\sl Nuclear Shell Theory} (New York: Academic Press).
\\[0.5ex]
Shayegan, M., Jo, J., Suen, Y. W., Santos, M., and Goldman, V. J., 1990,
{\sl Phys. Rev. Lett.}, {\bf65}, 2916.
\\[0.5ex]
Sitko, P., Yi, S. N., Yi, K.-S., and Quinn, J. J., 1996,
{\sl Phys. Rev. Lett.}, {\bf76}, 3396.
\\[0.5ex]
Sitko, P., Yi, K.-S., and Quinn, J. J., 1997,
{\sl Phys. Rev. B}, {\bf56}, 12~417.
\\[0.5ex]
Tsui, D. C., St\"ormer, H. L., and Gossard, A. C., 1982,
{\sl Phys. Rev. Lett.}, {\bf48}, 1559.
\\[0.5ex]
Willet R., Eisenstein, J. P., St\"ormer, H. L., 
Tsui, D. C., Gossard, A. C., and English, J. H., 1987,
{\sl Phys. Rev. Lett.}, {\bf59}, 1776.
\\[0.5ex]
W\'ojs, A., and Quinn, J. J., 1998a,
{\sl Physica E}, {\bf3}, 181.
\\[0.5ex]
W\'ojs, A., and Quinn, J. J., 1998b,
{\sl Solid State Commun.}, {\bf108}, 493.
\\[0.5ex]
W\'ojs, A., Hawrylak, P., and Quinn, J. J., 1998c,
{\sl Physica B}, {\bf256}-{\bf258}, 490.
\\[0.5ex]
W\'ojs, A., and Quinn, J. J., 1999a,
{\sl Solid State Commun.}, {\bf110}, 45.
\\[0.5ex]
W\'ojs, A., Hawrylak, P., and Quinn, J. J., 1999b,
{\sl Phys. Rev. B}, {\bf60}, 11~661.
\\[0.5ex]
W\'ojs, A., Szlufarska, I., Yi, K.-S., and Quinn, J. J., 1999c,
{\sl Phys. Rev. B}, {\bf60}, R11~273.
\\[0.5ex]
W\'ojs, A., and Quinn, J. J., 2000,
{\sl Phys. Rev. B}, {\bf61}, 2846.
\\[0.5ex]
Wu, T. T., and Yang, C. N., 1976,
{\sl Nucl. Phys. B}, {\bf107}, 365.
\\[0.5ex]
Yi, K.-S., Sitko, P., Khurana, A., and Quinn, J. J., 1996,
{\sl Phys. Rev. B}, {\bf54}, 16~432.
\end{raggedright}

\end{document}